\PassOptionsToPackage{unicode}{hyperref}
\PassOptionsToPackage{hyphens}{url}
\PassOptionsToPackage{dvipsnames,svgnames,x11names}{xcolor}
\documentclass[
  12pt]{article}
\usepackage{xcolor}
\usepackage{amsmath,amssymb}
\usepackage{iftex}
\ifPDFTeX
  \usepackage[T1]{fontenc}
  \usepackage[utf8]{inputenc}
  \usepackage{textcomp} 
\else 
  \usepackage{unicode-math} 
  \defaultfontfeatures{Scale=MatchLowercase}
  \defaultfontfeatures[\rmfamily]{Ligatures=TeX,Scale=1}
\fi
\usepackage{lmodern}
\ifPDFTeX\else
\fi
\IfFileExists{upquote.sty}{\usepackage{upquote}}{}
\IfFileExists{microtype.sty}{
  \usepackage[]{microtype}
  \UseMicrotypeSet[protrusion]{basicmath} 
}{}
\makeatletter
\@ifundefined{KOMAClassName}{
  \IfFileExists{parskip.sty}{%
    \usepackage{parskip}
  }{
    \setlength{\parindent}{0pt}
    \setlength{\parskip}{6pt plus 2pt minus 1pt}}
}{
  \KOMAoptions{parskip=half}}
\makeatother
\makeatletter
\ifx\paragraph\undefined\else
  \let\oldparagraph\paragraph
  \renewcommand{\paragraph}{
    \@ifstar
      \xxxParagraphStar
      \xxxParagraphNoStar
  }
  \newcommand{\xxxParagraphStar}[1]{\oldparagraph*{#1}\mbox{}}
  \newcommand{\xxxParagraphNoStar}[1]{\oldparagraph{#1}\mbox{}}
\fi
\ifx\subparagraph\undefined\else
  \let\oldsubparagraph\subparagraph
  \renewcommand{\subparagraph}{
    \@ifstar
      \xxxSubParagraphStar
      \xxxSubParagraphNoStar
  }
  \newcommand{\xxxSubParagraphStar}[1]{\oldsubparagraph*{#1}\mbox{}}
  \newcommand{\xxxSubParagraphNoStar}[1]{\oldsubparagraph{#1}\mbox{}}
\fi
\makeatother

\usepackage{longtable,booktabs,array}
\usepackage{calc} 
\usepackage{etoolbox}
\makeatletter
\patchcmd\longtable{\par}{\if@noskipsec\mbox{}\fi\par}{}{}
\makeatother
\IfFileExists{footnotehyper.sty}{\usepackage{footnotehyper}}{\usepackage{footnote}}
\makesavenoteenv{longtable}
\usepackage{graphicx}
\makeatletter
\newsavebox\pandoc@box
\newcommand*\pandocbounded[1]{
  \sbox\pandoc@box{#1}%
  \Gscale@div\@tempa{\textheight}{\dimexpr\ht\pandoc@box+\dp\pandoc@box\relax}%
  \Gscale@div\@tempb{\linewidth}{\wd\pandoc@box}%
  \ifdim\@tempb\p@<\@tempa\p@\let\@tempa\@tempb\fi
  \ifdim\@tempa\p@<\p@\scalebox{\@tempa}{\usebox\pandoc@box}%
  \else\usebox{\pandoc@box}%
  \fi%
}
\def\fps@figure{htbp}
\makeatother

\usepackage[]{natbib}
\usepackage{siunitx}
\usepackage{amsmath}
\usepackage{array, makecell}
\usepackage[table]{xcolor}
\usepackage{algorithm}
\usepackage{algorithmic}

\makeatletter
\def\@seccntformat#1{\@ifundefined{#1@cntformat}%
   {\csname the#1\endcsname\space}
   {\csname #1@cntformat\endcsname}}
\newcommand\section@cntformat{\thesection.\space} 
\makeatother

\makeatletter
\@ifpackageloaded{caption}{}{\usepackage{caption}}
\AtBeginDocument{%
\ifdefined\contentsname
  \renewcommand*\contentsname{Table of contents}
\else
  \newcommand\contentsname{Table of contents}
\fi
\ifdefined\listfigurename
  \renewcommand*\listfigurename{List of Figures}
\else
  \newcommand\listfigurename{List of Figures}
\fi
\ifdefined\listtablename
  \renewcommand*\listtablename{List of Tables}
\else
  \newcommand\listtablename{List of Tables}
\fi
\ifdefined\figurename
  \renewcommand*\figurename{Figure}
\else
  \newcommand\figurename{Figure}
\fi
\ifdefined\tablename
  \renewcommand*\tablename{Table}
\else
  \newcommand\tablename{Table}
\fi
}
\@ifpackageloaded{float}{}{\usepackage{float}}
\floatstyle{ruled}
\@ifundefined{c@chapter}{\newfloat{codelisting}{h}{lop}}{\newfloat{codelisting}{h}{lop}[chapter]}
\floatname{codelisting}{Listing}

\makeatother
\makeatletter
\@ifpackageloaded{caption}{}{\usepackage{caption}}
\@ifpackageloaded{subcaption}{}{\usepackage{subcaption}}
\makeatother
\usepackage{bookmark}
\IfFileExists{xurl.sty}{\usepackage{xurl}}{} 
\hypersetup{
  pdftitle={Penalized Fair Regression for Multiple Groups in Chronic Kidney Disease},
  pdfauthor={Carter H. Nakamoto; Lucia Lushi Chen; Agata Foryciarz; Sherri Rose},
  pdfkeywords={algorithmic bias, chronic kidney disease, fair
regression, penalized regression},
  colorlinks=true,
  linkcolor={blue},
  filecolor={Maroon},
  citecolor={Blue},
  urlcolor={Blue},
  pdfcreator={LaTeX via pandoc}}

\begin{document}

\def\spacingset#1{\renewcommand{\baselinestretch}%
{#1}\small\normalsize} \spacingset{1}


\title{\bf Penalized Fair Regression for Multiple Groups in Chronic
Kidney Disease}
\author{
Carter H. Nakamoto, cnaks21@stanford.edu\\
Department of Health Policy, Stanford University,\\ 615 Crothers Way
Encina Commons MC6019 Stanford, CA 94305, USA\\
Lucia Lushi Chen, \\
Psyfy Inc.\\
Agata Foryciarz, \\
Department of Epidemiology and Population Health, Stanford University\\
Sherri Rose, \\
Department of Health Policy, Stanford University\\
}
\maketitle

\bigskip
\begin{center}
\textbf{Short Running Head:} Penalized Fair Regression for Multiple Groups
\end{center}

\bigskip
\begin{abstract}
Fair regression methods have the potential to mitigate societal bias
concerns in health care, but there has been little work on penalized
fair regression when multiple groups experience such bias. We propose a
general regression framework that addresses this gap with unfairness
penalties for multiple groups. Our approach is demonstrated for binary
outcomes with true positive rate disparity penalties and can be
efficiently implemented through reduction to a cost-sensitive
classification problem. We additionally introduce novel score functions
for automatically selecting penalty weights. Our penalized fair
regression methods are empirically studied in simulations, where they
achieve a fairness-accuracy frontier beyond that of existing comparison
methods. Finally, we apply these methods to a national multi-site
primary care study of chronic kidney disease to develop a fair
classifier for end-stage renal disease. There we find substantial
improvements in fairness for multiple race and ethnicity groups who
experience societal bias in the health care system without any
appreciable loss in overall fit.
\end{abstract}

\noindent%
\vfill

\newpage
\spacingset{1.9} 

\section{INTRODUCTION}\label{introduction}

Fair regression methods modify the loss function to encode fairness
measures. These loss functions typically include a single penalty or
constraint based on a measure of unfairness for one group
\citep{calders_controlling_2013, berk_convex_2017}. Subsequent work has
developed technical methods for solving single group penalized or
constrained fair regression problems
\citep[e.g.,][]{dwork_decoupled_2018} and tailored them for health
applications \citep{zink_fair_2020}.

We develop penalized fair regression methods for multiple groups in an
application to classify whether high-risk individuals with chronic
kidney disease (CKD) will progress to end stage renal disease (ESRD).
CKD is prevalent, affecting an estimated 14\% of the US adult population
\citep{cdc_chronic_2025}. In severe cases, CKD leads to ESRD, requiring
either dialysis or kidney transplant, but progress can be delayed
through treatment and management. Black, Hispanic, Native American, and
Asian patients all have higher incidence and prevalence of ESRD than
white patients, and they have worse kidney function, on average, than
white patients when initiating treatment for ESRD
\citep{johansen_us_2025}. Identifying patients at risk for ESRD could
help these patients engage in more intensive treatment and slow CKD
progression, and fair regression is a potential approach to partially
address disparities in progression to ESRD. Another key algorithm for
defining CKD progression, the estimated glomerular filtration rate
(eGFR) equation, was recently revised to remove race adjustment, so CKD
may be a context where it is especially appropriate to use a fair
regression method that can increase fairness without race adjustment
\citep{cusick_algorithmic_2024}.

Although there is a mathematical equivalence between penalized and
constrained methods such that each penalized regression has a
corresponding constrained regression with the same solution, most
methods formulate their fair regressions as constrained optimization
problems. This overlooks that penalty-based approaches have several
advantages over constraint-based approaches. For one, the approaches are
conceptually distinct. A constraint is a user-imposed maximum level of
unfairness, while a penalty is a user-imposed tradeoff between accuracy
and fairness. These conceptions are appropriate for different
applications and their social and regulatory contexts. The two
approaches may also require divergent computational implementations.
Much of the constrained optimization literature develops original
solution algorithms
\citep{kearns_preventing_2018, komiyama_nonconvex_2018}, whereas many
existing algorithms can minimize penalized loss functions under
different conditions, including gradient descent and Nelder-Mead
\citep{nelder_simplex_1965, loshchilov_decoupled_2019}. Outside of the
fairness literature, penalized methods are common in statistics
applications
\citep{hoerl_ridge_1970, tibshirani_regression_1996, zou_regularization_2005}
and fair penalty methods can benefit from advances established in those
applications.

Fair regression has the potential to reduce the impact of societal bias
in health care. However, only a small number of fair regression works
have examined realistic use cases in health care
\citep{zink_fair_2020, mcguire_improving_2021, pfohl_empirical_2021, pfohl_net_2022, wang_enhancing_2025}.
Health care applications have complex ethical considerations, many of
which are unique due to the complexity of medical systems
\citep{chen_ethical_2021, chen_algorithmic_2023}. For example, questions
about how to handle certain patient characteristics in clinical
algorithms, such as race, are nuanced such that not all fair regression
approaches may be appropriate \citep{cusick_algorithmic_2024}. Another
major limitation of most fair regression approaches is that they focus
on fairness with respect to a single group when multiple groups
typically face unfairness in real world settings
\citep{chen_ethical_2021, liu_practical_2022, wang_enhancing_2025}. The
joint consideration of multiple groups in fair regression is
non-trivial, with challenges related to the optimal weighting of the
groups as well as the additional computational complexity
\citep[e.g.,][]{kearns_preventing_2018}. We refer here to multiple
groups rather than sensitive attributes to avoid the particular social
and sometimes legal connotations of the latter, reflecting the breadth
of possible groups for which these methods may be appropriate. In health
settings, groups might be defined by race, sex, or additional
attributes.

While limited compared to single group methods, there is literature on
fair constrained regression for multiple groups establishing frameworks
and efficient solutions
\citep[e.g.,][]{kearns_preventing_2018, wang_enhancing_2025}. Of
particular note, \citet{agarwal_reductions_2018} proposed a method that
reduces fair binary classification to a sequence of cost-sensitive
classification problems, enabling the use of efficient off-the-shelf
classifiers for fair constrained regression. This approach is flexible
across many fairness metrics and was later extended, including to
continuous outcomes \citep{agarwal_fair_2019} and to non-linear but
convex constraints, including demographic parity and equalized odds, and
convex or fractional-convex losses \citep{narasimhan_learning_2018}.
These methods, like most constrained optimization methods, involve a
min--max saddle-point game between a predictor and an adversary
enforcing fairness constraints, often leading to non-convex objectives,
instability, substantial hyperparameter sensitivity, and high
computational costs, especially with imbalanced group sizes.

In contrast to constrained regression, we have identified even fewer
papers that implement penalized fair regression with respect to multiple
groups, highlighting the lesser focus on this formulation of the
estimation problem. \citet{perez-suay_fair_2017} used the Hilbert
Schmidt independence criterion for fairness penalties. This method only
accommodates one measure of fairness, demographic parity, which is
satisfied when estimated outcomes are independent of groups. A ridge
penalty method that builds on \citet{komiyama_nonconvex_2018} to enforce
demographic parity was developed in \citet{scutari_achieving_2022}.
Their approach includes group identifiers as predictors, which will not
be appropriate for some estimation problems. They also proposed an
extension to other fairness metrics, including equalized odds. Moreover,
the focus of these first two papers on demographic parity restricts
applications in health care, where pressing fairness concern may differ
from and be incompatible with demographic parity
\citep{seyyed-kalantari_underdiagnosis_2021, chen_algorithmic_2023}.
\citet{pfohl_empirical_2021} and \citet{pfohl_net_2022} implement
penalized fair regression with a single penalty term that measures
unfairness with respect to a single categorical group, capturing a
subset of multiple group fair regression problems. Lastly,
\citet{liu_practical_2022} introduced a fairness measure built on
cross-covariance operators on reproducing kernel Hilbert Spaces which
were used to construct unfairness penalties. While the approach is
flexible with respect to fairness measures, its complexity presents an
implementation barrier, and no code is available.

We propose a general framework for regression with unfairness penalties
for multiple groups. Logistic regression is implemented as a
demonstration estimator, and true positive rate disparity is used as the
demonstration unfairness measure, although our framework is broad and
allows for other choices. The penalized regression problem is solved
efficiently through conversion into a cost-sensitive classification
problem and, in turn, into a weighted classification problem. Optimal
penalties for each group's unfairness measure are automatically selected
given a user-specified tradeoff between accuracy and fairness via a
random search and score function. We apply this method to the problem of
CKD progression classification, penalizing unfairness for Black,
Hispanic non-white, and Asian patients.

This work contributes to several understudied areas of the literature.
The joint consideration of fairness with respect to multiple groups
remains novel, and we additionally derive multiple ways of
mathematically synthesizing fairness information across groups. Further,
we introduce a method for choosing a different penalty for each group
automatically based on a preferred fairness-accuracy tradeoff. The
application of fair regression to health care with the contextually
appropriate fairness metric of true positive rate disparity is also new.

In Section~\ref{sec-rp}, we describe the research problem, establishing
the variables and defining fairness for this problem. In
Section~\ref{sec-methods}, we introduce the penalized regression
approach. We define the loss function and describe how we select
penalties using a random search with novel score functions. In
Section~\ref{sec-sim}, we conduct a simulation study to examine our
method's performance against other multi-group fair regression methods
with varied data distributions and score functions. In
Section~\ref{sec-application}, we apply our methods to estimating CKD
progression while penalizing unfairness for Black, Hispanic non-white,
and Asian patients and present performance metrics for the resulting
classifiers. We conclude in Section~\ref{sec-discussion}.

\section{RESEARCH PROBLEM}\label{sec-rp}

Given a finite set of groups \(G\), we seek to classify an outcome
\(Y=\{0,1\}\) as a function of input variables \(X\) in a manner that is
fair with respect to each group \(g \in G\), where each \(g\) has
\(n_g\) individuals. We will also define a reference group \(r\), a set
of individuals pre-specified to be an appropriate comparison.
Individuals may be part of multiple groups, including the reference
group, as appropriate for the estimation task, and some individuals may
not belong to any group, including the reference group. Let
\(\mathbb{I}_{g}\) be a random variable indicating membership in group
\(g\) and \(\mathbb{I}_{r}\) be a random variable indicating membership
in reference group \(r\). Our observed data contain \(n\) individuals,
indexed by \(i\).

In this article, we use true positive rate equality, based on equal
opportunity \citep{hardt_equality_2016}, as our measure of fairness,
where equality holds if the estimator true positive rates are the same
for members of group \(g\) and members of reference group \(r\). A focus
on true positive rate fairness (or, equivalently, false negative rate
fairness) is appropriate for the problem of low rates of screening and
diagnosis among high-risk patients in certain groups
\citep{seyyed-kalantari_underdiagnosis_2021}. This is especially true
when the benefit of a diagnosis is much greater than the potential harm,
as is the case for our CKD progression application, because this implies
that false negatives are much more costly than false positives and
should be the primary fairness concern. We can express different
versions of this fairness measure for estimated outcomes and for
estimated probabilities. The estimated outcome formulation is more
interpretable. However, the estimated probability formulation is
differentiable, so it can be easily minimized, and it does not depend on
a particular classification threshold.

Let \(j\) index individuals in group \(g\), \(k\) index individuals in
reference group \(r\), and \(U\) be a measure of unfairness. For
estimated outcomes, the equal opportunity unfairness measure \(U^{EO}\)
for a group is the difference between reference group and group true
positive rates:

\begin{equation}\phantomsection\label{eq-fair-yhat}{
U^{EO} = \frac{\sum_{k \in r} Y_k \hat Y_k}{\sum_{k \in r} Y_k} - \frac{\sum_{j\in g}Y_j \hat Y_j}{\sum_{j\in g}Y_j},
}\end{equation} where \(\hat{Y} = \mathbb{I}(\hat{P} > \zeta)\), for a
classification threshold of \(\zeta\), and \(\hat{P}=P(Y=1 | X)\). For
estimated probabilities, we substitute \(\hat{P}\) for \(\hat{Y}\) as
has been done in prior literature \citep{wang_enhancing_2025}. This
version of the unfairness measure, which we call \(U^{EOP}\), is given
by: \begin{equation}\phantomsection\label{eq-fair-phat}{
U^{EOP} = \frac{\sum_{k \in r} Y_k \hat P_k}{\sum_{k \in r} Y_k} - \frac{\sum_{j\in g}Y_j \hat P_j}{\sum_{j\in g}Y_j}.
}\end{equation}

There is no consensus in the literature on how to evaluate fairness for
multiple groups jointly, as many past fair regression works characterize
unfairness for each group individually
\citep{agarwal_reductions_2018, komiyama_nonconvex_2018}. Given the
unfairness metrics in Equations \eqref{eq-fair-yhat} and
\eqref{eq-fair-phat} for a given group and reference, which we will
denote \(U_g\), we study three approaches to aggregating across groups.
First, we propose a group population-weighted average:
\(U^{PW} = \sum_{g \in G} n_g U_g / \sum_{g \in G} n_g\). Second, we
propose a group-weighted average, giving equal weight to all groups:
\(U^{GW} = \sum_{g \in G} U_g / \sum_{g \in G} 1\). Third, we study a
maximum across all group-level unfairness measures, based on
\citet{ghosh_characterizing_2021}: \(U^{Max} = \max_{g \in G} U_g\).

\section{METHODS}\label{sec-methods}

Our novel penalized fair regression approach generalizes to various
fairness metrics, outcome variable types, and estimator architectures.
We focus our main demonstration on logistic regression with true
positive rate disparity penalties. Web Appendix A contains loss
functions for continuous outcomes with mean residual difference as the
unfairness metric.

\subsection{Regression with Unfairness Penalties}\label{sec-penloss}

The penalty term for each group \(g\) is the product of group-specific
penalty weight \(\lambda_g\) and group unfairness \(U_g\). For logistic
regression, the penalized loss combines binary cross-entropy loss with
the unfairness measure in Equation \eqref{eq-fair-phat} and penalty
weights \(\lambda_g\). We also multiply the group unfairness penalty
terms by group size \(n_g\). The resulting loss function is \[
L = - \sum_{i=1}^n \{ Y_i\ln(\hat P_i) + (1-Y_i)\ln(1 - \hat P_i) \} + \sum_{g \in G} \lambda_g n_g \left(\frac{\sum_{k\in r}Y_k\hat P_k}{\sum_{k\in r}Y_k} - \frac{\sum_{j\in g}Y_j\hat P_j}{\sum_{j\in g}Y_j} \right). 
\] When \(\lambda_g = 0 \forall g \in G\), the loss function becomes
binary cross-entropy loss, and for extremely large values of
\(\lambda_g \forall g \in G\), the loss function is minimized with the
estimates \(\hat P_k = 0 \forall k \in r\) and
\(\hat P_j = 1 \forall j \in g \forall g \in G\). This loss function is
convex but has no closed form minimizing solution. We employ an
efficient optimization method based on the reductions approach of
\citet{agarwal_reductions_2018}. In this approach, we use cost
minimization loss to reconceptualize the penalized fair regression as a
cost-sensitive classification problem, which can be implemented as a
weighted non-cost-sensitive classification problem with weights \[
W_i = |C_i^0 - C_i^1| = \left|Y_i\left \{ 2 - \sum_{g \in G}\lambda_g n_g \left( \frac{\mathbb{I}_{ri}}{\sum_{k\in r}Y_k } - \frac{\mathbb{I}_{gi}}{\sum_{j\in g}Y_j}  \right) \right\} - 1\right|,
\] and modified outcomes \(Y_i' = \mathbb{I}(C_i^0 > C_i^1)\). We can
then perform the penalized fair regression with a wide array of standard
classifiers (e.g., logistic regression, decision trees) compatible with
individual-level sample weighting. Detailed derivations are in Web
Appendix B.

\subsection{Selecting Penalty Coefficients}\label{sec-pen}

Next, we identify the appropriate unfairness penalty weights
\(\lambda_g \forall g \in G\). We propose an original score function
that is the weighted average of rescaled measures of accuracy and
fairness, with a user-specified weight \(\alpha\), which can be thought
of as a fairness score weight, to determine the relative contribution of
accuracy and fairness for the score. The accuracy term is a
transformation of accuracy rescaled such that it is \(1\) at a
reasonable lower bound for accuracy and \(0\) at a reasonable upper
bound. The basis for the accuracy term is the share of units for which
the estimator's classification matches the outcome:
\(A = \frac1n \sum_{i=1}^n \mathbb{I}(Y_i= \hat Y_{i})\). We set the
reasonable upper bound to be the unpenalized estimator's accuracy
\(A^{UP}\) and the reasonable lower bound at \(0.5\). The corresponding
rescaled accuracy term is \(A^{RS} = (A^{UP} - A)/(A^{UP} - 0.5)\).

Similarly, the fairness term is a transformation of a fairness metric
rescaled to be \(1\) at a reasonable lower bound for fairness and \(0\)
at a reasonable upper bound. The basis for the fairness term is true
positive rate disparity, measured as in Equation \eqref{eq-fair-yhat}.
For each group, we take the maximum of the unfairness term and \(0\),
considering unfairness to be the extent to which the true positive rate
of group \(g\) is lower than that of the reference group \(r\). We
synthesize across groups using either a population-weighted average,
group-weighted average, or maximum, and rescale by dividing by the
analagous measure for the standard unpenalized estimator. Let
\(\hat Y^{UP}\) denote the unpenalized estimator's outcome estimate. The
rescaled unfairness term in the population-weighted specification is \[
U^{RS} = \frac{\sum_{g \in G}n_g \max \left( U_g, 0\right)}{\sum_{g \in G}n_g \max \left( U^{UP}_g,0\right)} = 
\frac{\sum_{g \in G}n_g \max \left( \frac{\sum_{k\in r}Y_k\hat Y_k}{\sum_{k\in r}Y_k} - \frac{\sum_{j\in g}Y_j \hat Y_{j}}{\sum_{j\in g}Y_j}, 0\right)}{\sum_{g \in G}n_g \max \left(\frac{\sum_{k\in r}Y_k\hat Y^{UP}_k}{\sum_{k\in r}Y_k} - \frac{\sum_{j\in g}Y_j\hat Y^{UP}_j}{\sum_{j\in g}Y_j},0\right)}.
\] The above measure equals zero if the candidate estimator achieves
\(0\) unfairness for all groups and to \(1\) if the candidate estimator
has the same multi-group unfairness as the unpenalized estimator. In the
group-weighted specification, the rescaled unfairness term is
\(\sum_{g \in G} \max \left( U_g, 0\right) / \sum_{g \in G} \max \left( U^{UP,g},0\right),\)
and in the maximum specification, the rescaled unfairness term is
\(\max_{g \in G} \left \{ \max \left( U_g, 0\right) \right \} / \max_{g \in G} \left \{\max \left( U^{UP,g},0\right) \right \}.\)

Finally, the score is the negative weighted average of rescaled accuracy
and rescaled fairness with a user-specified weight \(\alpha \in [0,1]\):
\(-s = \alpha U^{RS} + (1-\alpha) A^{RS}\). The negative is used so that
maximizing the score function corresponds to higher accuracy and
fairness. For \(\alpha = 0\), maximizing the score function is
equivalent to maximizing accuracy with no fairness concerns as in an
unpenalized regression estimator. For \(\alpha = 1\), maximizing this
score function is equivalent to minimizing true positive rate
disparities without regard for accuracy.

Our penalty weight search is a random search, which yields higher
maximum scores more efficiently than a grid search
\citep{bergstra_random_2012}. A number of sets of candidate values for
\(\lambda_g\) are drawn, and an estimator is fit for each set. The
highest scoring penalty weight set is selected. All group penalty weight
candidates are sampled from identical independent log uniform
distributions. We favor this approach over greedy sequential approaches
because it allows for optimization of multiple score functions
simultaneously. The breadth of the penalty weight candidate search space
and number of candidate draws will vary for each penalized fair
regression problem, but we include some empirical demonstrations of
different choices in our simulation study. We implement five-fold cross
validation to evaluate the random penalty weight candidates, fitting an
estimator on the \(V-1\) folds, calculating the score on the \(V\)th
fold, and selecting the highest scoring penalty weight set.

\subsection{Computational
Implementation}\label{computational-implementation}

The proposed penalized regression estimator is implemented using a
custom \texttt{sklearn} estimator with the \texttt{BaseEstimator}
function, primarily to encode the construction of penalty-based sample
weights for the reduction and implement the novel score function. The
logistic regression itself is fit with the base implementation from
\texttt{sklearn}. Random search steps in the penalty weight selection
process are implemented with cross-validation facilitated by
\texttt{sklearn}'s \texttt{KFold} but without using \texttt{sklearn}'s
\texttt{RandomSearchCV} to accommodate multiple complex score functions.
Score functions have original implementations to accommodate the
inclusion of multiple groups, reference groups, and unpenalized
comparison estimators. We do not incorporate methods to adjust for class
imbalance due to inconsistent to poor performance of these tools in
practical applications
\citep{blagus_smote_2013, adhikari_revisiting_2021}. This algorithm is
available in Web Appendix B. Project code is in a public GitHub
repository
(\href{https://github.com/StanfordHPDS/penalized_multiple_groups}{github.com/StanfordHPDS/penalized\_multiple\_groups}).

\section{SIMULATION STUDY}\label{sec-sim}

We conducted a simulation study to demonstrate the performance of our
proposed method and comparison methods across different data
distributions. In the simulation, the groups have separate data
generating processes and true positive rate disparities with respect to
the reference group. Specifically, we generated a population of
1,000,000 per simulation with three groups, with many individuals
belonging to multiple groups. Simulation setting 1 had all moderately
sized groups, simulation setting 2 had one small group, and all groups
were small in simulation setting 3. The data generating process is
available in Web Appendix C.

\subsection{Simulation Implementation}\label{sec-simmethods}

We explored the performance of various score functions using each of the
group synthesis mechanisms in Section~\ref{sec-pen} with fairness score
weights \(\alpha \in \{0.1, 0.2, \dots, 0.9, 0.99 \}\). Additionally, we
varied the number of candidate penalty weight set draws
\([10, 20, \dots, 100, 200, 400]\), where each candidate penalty weight
set consists of one value of \(\lambda_g\) for each group \(g \in G\).
Lastly, we studied robustness to estimator misspecification by omitting
one of the nine variables from our main terms regressions. Comparison
estimators were baseline unpenalized regression and the fair reductions
constrained regression method from \citet{agarwal_reductions_2018} as
implemented in the \texttt{fairlearn} python package. The exponentiated
gradient algorithm was implemented with allowed fairness constraint
violations of \(0.1\) and \(0.01\), and the grid search algorithm was
used with constraint weights of \(0.25\), \(0.5\), and \(0.75\).
Otherwise, package default hyperparameters were used.

Our simulations featured 1000 replications of the full data generation
and random penalty weight search process for each simulation setting and
number of candidate penalty weight sets per random search. Penalty
weight candidates were all sampled from \(\text{LogUnif}(-3,1)\). This
distribution was an ad hoc empirical choice based on the range of values
in our simulation. Five-fold cross validation was used for the random
penalty weight search. In each random search, we simultaneously
maximized all score function variations.

\subsection{Simulation Results}\label{simulation-results}

\begin{figure}

\centering{

\pandocbounded{\includegraphics[keepaspectratio]{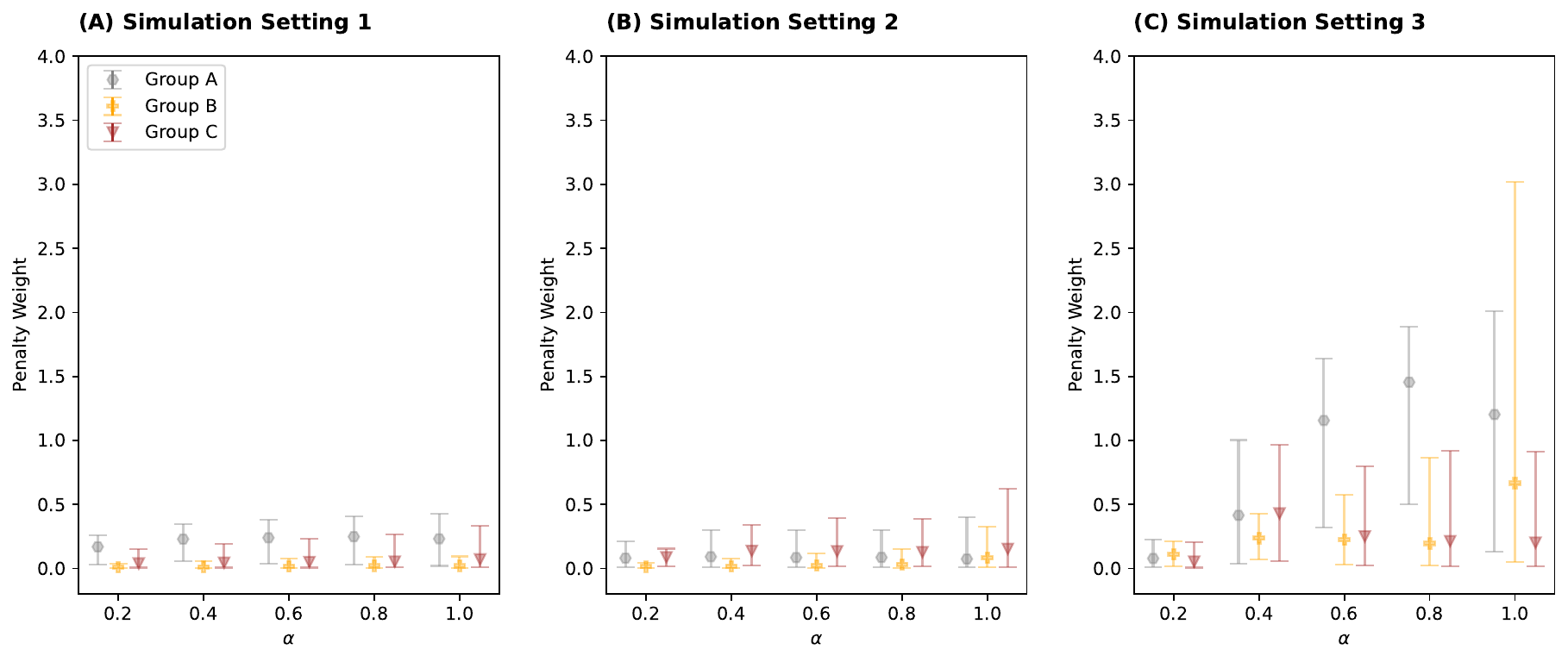}}

}

\caption{\label{fig-1}\textbf{Penalty Weights for Each Group by Fairness Score Weight.}
Bars represent the empirical 25th and 75th percentiles of replications.
Simulations with 40 draws of candidate penalty weight sets are
represented.}

\end{figure}

Our novel penalized regression methods were evaluated across several key
dimensions: behavior of unfairness penalty weights by fairness score
weights, understanding penalty weights across various evaluation
measures, effect of changes in multi-group unfairness synthesis
function, and the tradeoff between fairness and accuracy. The simulation
settings varied in group sizes, as detailed in Web Appendix C, Table 3.
First, we characterized the behavior of unfairness penalty weights under
the averaged score functions. The random search yielded different
unfairness penalty weights for the three groups across the
user-specified fairness score weights (Figure~\ref{fig-1}). Unfairness
penalty weights increased, on average, as the fairness score weight
\(\alpha\) increased. Unfairness penalty weights also varied across
groups and simulation settings.

\begin{figure}

\centering{

\pandocbounded{\includegraphics[keepaspectratio]{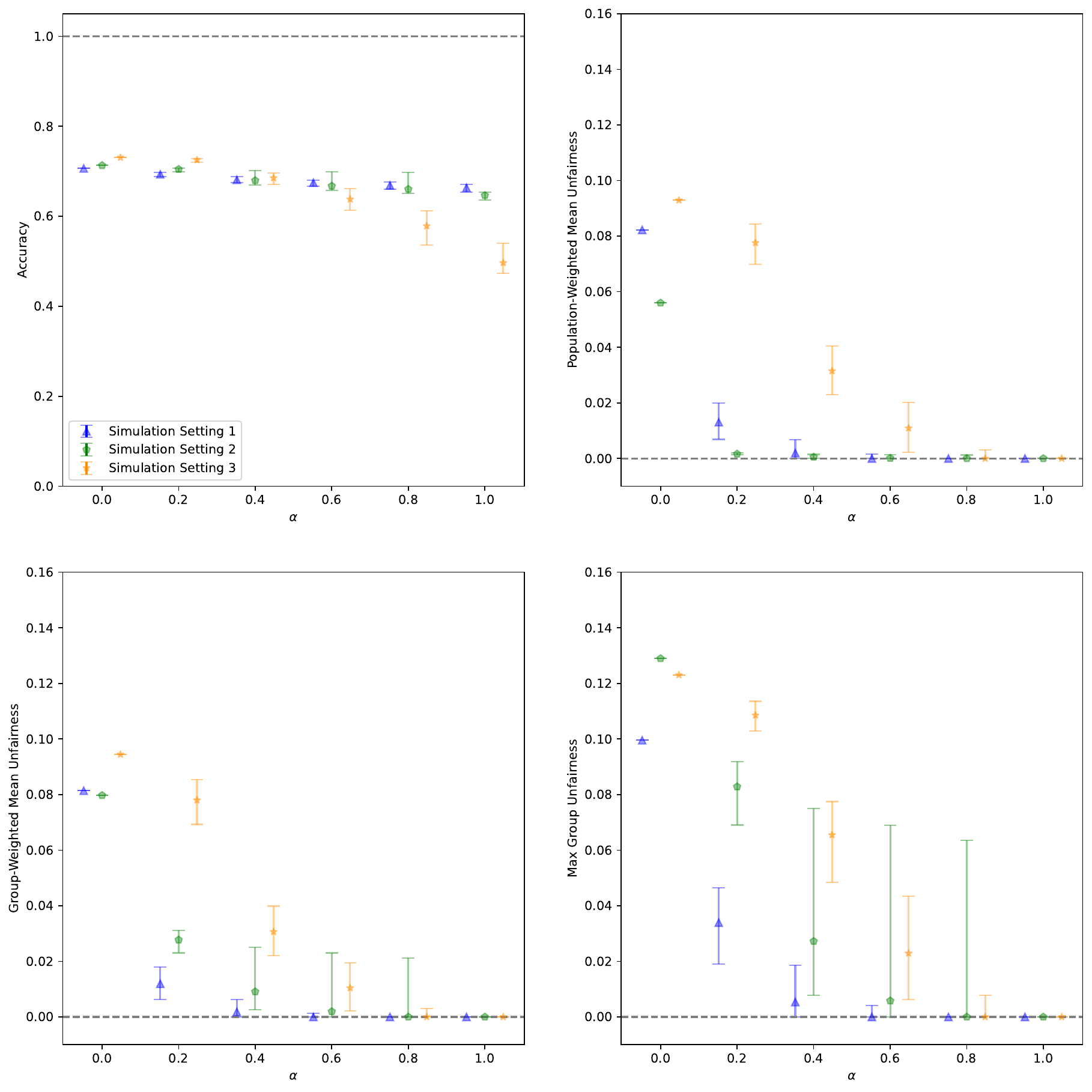}}

}

\caption{\label{fig-2}\textbf{Fair Regression Estimator Accuracy and Unfairness for Different Fairness Score Weights by Simulation Setting.}
Bars represent the empirical 25th and 75th percentiles of replications.
Simulations with 40 draws of candidate penalty weight sets are
represented.}

\end{figure}

The different unfairness penalty weights yielded different fairness and
accuracy measures (Figure~\ref{fig-2}). In general, for each simulation
setting, there was some range of user-specified fairness score weights
\(\alpha\) over which there were sharp decreases in unfairness with only
modest decreases in accuracy. For simulation setting 1, our penalized
method with \(\alpha = 0.5\) had an accuracy decrease of 3 percentage
points and a mean population-weighted unfairness decrease of 8
percentage points relative to the baseline estimator. Our penalized
method incurred steeper accuracy losses for simulation setting 3, where
\(\alpha = 0.5\) yielded an accuracy decrease of 7 percentage points and
a mean population-weighted unfairness decrease of 7 percentage points
relative to the baseline estimator.

\begin{figure}

\centering{

\pandocbounded{\includegraphics[keepaspectratio]{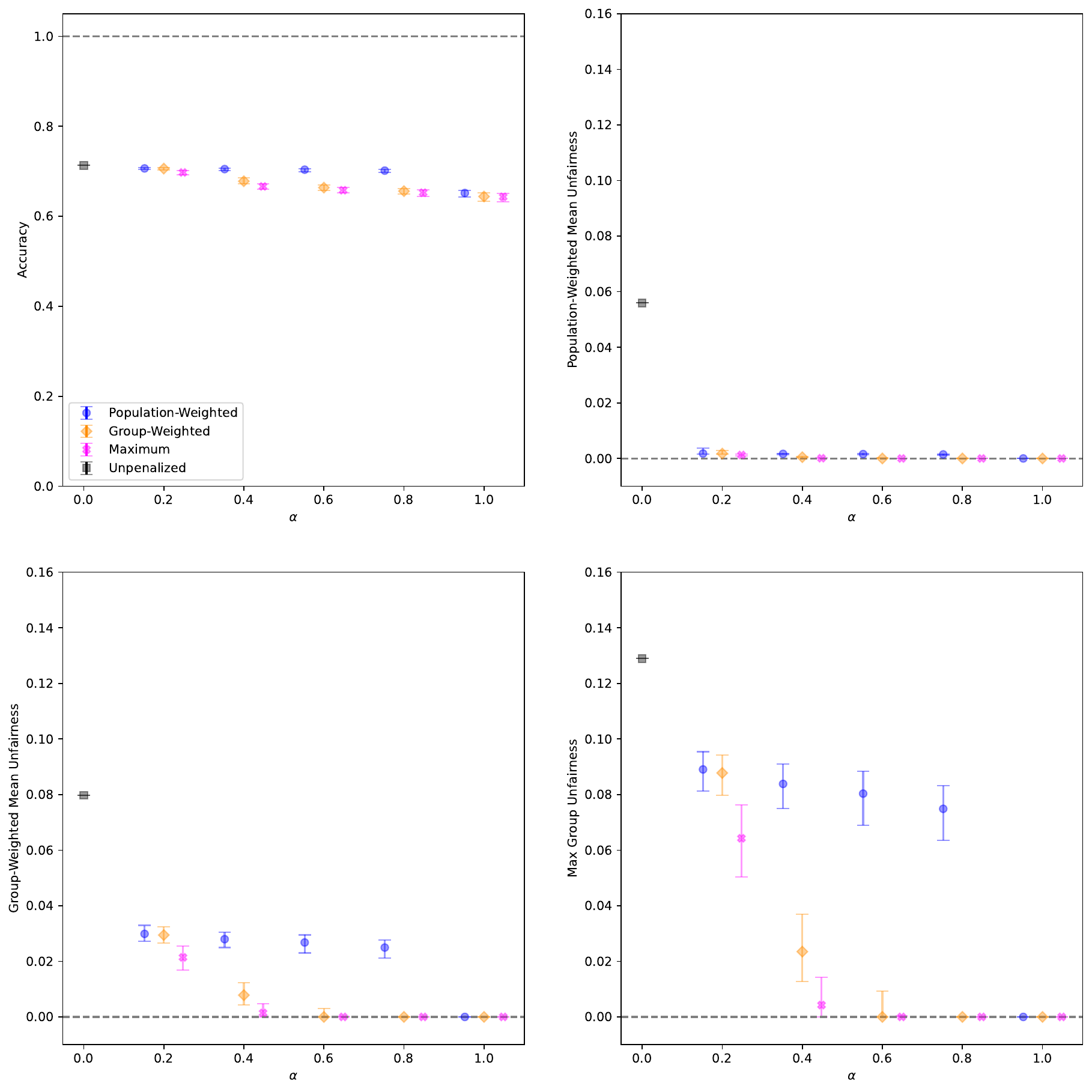}}

}

\caption{\label{fig-3}\textbf{Fair Regression Estimator Accuracy and Unfairness for Different Fairness Score Weights by Multi-Group Unfairness Synthesis Function, Simulation Setting 2.}
The subfigures correspond to performance measures for fitted estimators,
while colors correspond to how the fairness components of the score
functions were calculated. Bars represent the empirical 25th and 75th
percentiles of replications. Simulations with 40 draws of candidate
penalty weight sets are represented.}

\end{figure}

Next, we examine how accuracy and unfairness vary across multi-group
synthesis functions. Results only vary substantially by synthesis
function in simulation setting 2, which has one small group and two
larger groups (Figure~\ref{fig-3}). In this setting, score functions
with group-weighting or maximum synthesis yielded lower group-weighted
and maximum unfairness measures as well as lower accuracy. For fairness
score weight \(\alpha=0.5\), using maximum unfairness in score functions
yields 7 percentage points lower maximum unfairness and 4 percentage
points lower accuracy on average compared to using population-weighting.
In simulation settings 1 and 3, where groups are similar in size,
estimator performance is similar across multi-group synthesis functions
(Web Appendix C, Figures 7-8).

\begin{figure}

\centering{

\pandocbounded{\includegraphics[keepaspectratio]{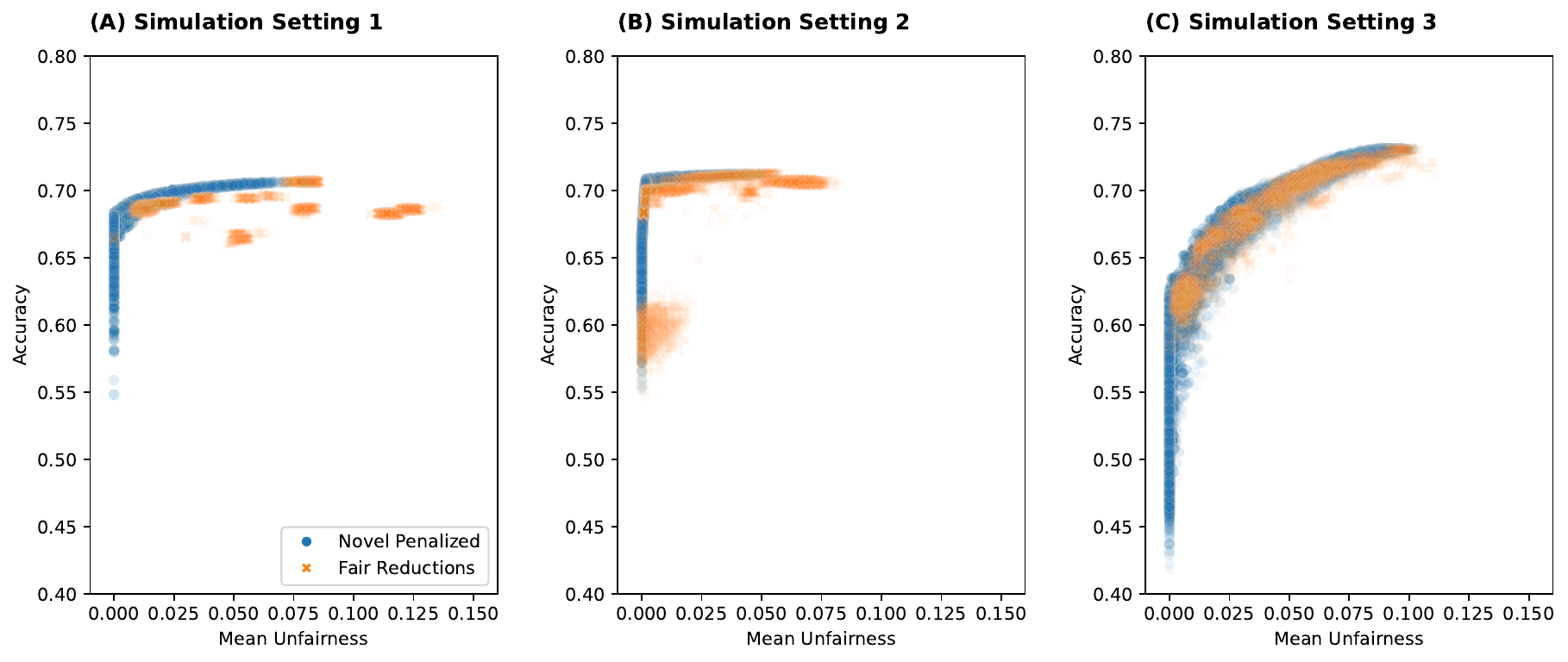}}

}

\caption{\label{fig-4}\textbf{Accuracy and Mean Fairness for Proposed and Comparison Fair Reductions Methods by Simulation Setting.}
Each plotted point corresponds to the result of a single simulation
under one of the score function variations or enumerated fair reductions
hyperparameters. Simulations with 40 draws of candidate penalty weight
sets are represented.}

\end{figure}

We compare the fairness and accuracy results of our penalized method to
the fair reductions constrained regression method (Figure~\ref{fig-4}).
This figure also shows the fairness-accuracy frontier, the set of
results for which there is no better performance in both fairness and
accuracy \citep{liang_algorithmic_2023}. The fairness-accuracy frontier
for our penalized method was beyond that of the comparison fair
reduction methods; the comparison method did not yield estimators that
outperformed our method's estimators in both mean fairness and accuracy.
In simulation settings 1 and 2, the penalized methods were clustered
around the fairness-accuracy frontier, while the comparison method had
more variance. For simulation setting 3, which featured smaller groups,
there was more variance across simulations for both methods, with more
replications far from the fairness-accuracy frontier.

Results with variation in the number of candidate penalty weight sets
and estimator misspecification are in Web Appendix C, Figures 5-6.
Accuracy and unfairness varied less as candidate penalty weight set
counts increased, and misspecified estimators generally had
fairness-accuracy frontiers shifted downward. Runtime data are available
in Web Appendix C. We detail our penalized fair regression methods with
an error rate disparity fairness metric, implementations with a random
forest based estimator, and further simulation studies in Web Appendix
D. We found similar patterns in the results as in our primary analyses,
with estimator performance on or near the fairness-accuracy frontier
(Web Appendix D, Figures 9-11).

\section{CHRONIC KIDNEY DISEASE APPLICATION}\label{sec-application}

\subsection{Data and Implementation}\label{data-and-implementation}

We applied our methods in the American Family Cohort (AFC) dataset
derived from the American Board of Family Medicine (ABFM) PRIME Registry
\citep{phillips_prime_2017, stanford_center_for_population_health_sciences_afc_2024}.
These primary care electronic health records from across the US contain
diagnosis and procedure codes as well as laboratory test and clinical
observation measurements. Using input variables measured in 2017-2018,
we predicted the occurrence of ESRD in 2019, defined as either an ESRD
diagnosis code or a calculated eGFR under 15. eGFR was calculated using
the CKD-EPI 2021 equation, which does not adjust for race, to more
accurately reflect kidney function \citep{inker_new_2021}.

Predictors were selected based on prior literature
\citep{segal_machine_2020, inoguchi_simplified_2022}. The physiologic
predictors were the means of diastolic and systolic blood pressure, the
maximum and mean values for hemoglobin A1C and billirubin, and the
minimum eGFR. Missingness in these predictors was accounted for by
including indicators of missingness as predictors. Indicators for
hypertension emergencies, type 2 diabetes, glomerulonephritis,
proteinuria, and pyelonephritis, constructed using diagnosis codes as
described in Web Appendix E, Table 4, were also included as predictors.
Finally, binary documented sex, year of birth, race, and ethnicity were
available in the AFC dataset. We included documented sex and age as
predictors. Race and ethnicity were used to define the groups of
interest for unfairness penalties and metrics, but we intentionally did
not include these variables as predictors. We studied fairness for three
race and ethnicity groups: Black, Hispanic non-white, and Asian. Some
patients were both Black and Hispanic non-white as well as Asian and
Hispanic non-white. White patients were the reference group. Two groups
facing CKD health disparities had small sample sizes (American Indian
and Alaska Native as well as Native Hawaiian and Pacific Islander) in
our dataset and were not included as groups for fairness optimization.

We restricted our sample to patients with diastolic and systolic blood
pressure measurements in 2017-2018 and to practices with at least one
ESRD patient in 2017-2019 for data quality. To target high-risk
susceptible patients, the sample was restricted to patients at least 35
years old with at least two CKD diagnosis codes, no documented ESRD
prior to 2019, and a maximum systolic blood pressure of at least 140
mmHg, indicating stage 2 hypertension, in 2017-2018
\citep{hsu_elevated_2005, cusick_balancing_2025}. Finally, we required
recorded values of either race or ethnicity. The resulting sample had
67901 patients, 2817 of whom developed ESRD in 2019. Detailed sample
characteristics are described in Web Appendix E, Table 5.

Our implementation deployed main terms penalized logistic regression. We
used a classification threshold of 0.15 for estimated ESRD probability,
informed by evidence that the benefits of increasing the intensity of
CKD treatment outweigh the costs and thus that false negatives are more
harmful than false positives, but other thresholds could also be valid
\citep{tisdale_cost-effectiveness_2022}. Penalty weight candidates were
sampled from the distribution \(\text{LogUnif}(-4,-1)\), which was
selected based on the range of values in our analysis. We used five-fold
cross validation to simultaneously identify the penalty weight sets that
maximized each score function variation described in
Section~\ref{sec-simmethods}. Comparison estimators were the same as in
our simulation with the algorithms and hyperparameters described in
Section~\ref{sec-sim}.

\subsection{Results}\label{results}

The baseline estimator had an accuracy of 0.95, a sensitivity of 0.37,
and a specificity of 0.97 (Table~\ref{tbl-baselogreg}). However, its
accuracy and sensitivity were lower for Black and Asian patients than
for the overall sample, while the opposite was true for white patients.
Accuracy was lower for Hispanic non-white patients compared to the
overall sample, but not sensitivity, although sensitivity was lower for
Hispanic non-white patients than for white patients. Hispanic non-white
patients had the highest ESRD prevalence among the groups we considered.

Under 30 different score function formulations (10 fairness score
weights with 3 multi-group synthesis functions), 5 distinct sets of
penalty weights maximized the score functions
(Table~\ref{tbl-esrdresmain}). In most cases, multiple score function
formulations yielded identical results. Overall, the penalized
estimators had 0-5 percentage point decreases in accuracy with 2-6
percentage point decreases in unfairness. The penalized regression
estimators with group-weighted fairness scores all had the same
performance. Compared to the baseline, these estimators had less than a
percentage point difference in accuracy, 1 percentage point higher
sensitivity, and 6 percentage points lower mean unfairness (an 85\%
reduction) while reducing unfairness for Black and Hispanic non-white
patients to 0. Thus, these estimators provided no appreciable loss in
overall fit with only improvements in fairness. However, across most
estimators, substantial unfairness remained for Asian patients, the
smallest group we considered in our data sample, comprising only 2\% of
patients. Some estimators using maximum unfairness as the group
synthesis mechanism achieved unfairness as low as 2-4 percentage points
for Asian patients, but at the cost of 6-11 percentage points of
sensitivity relative to the baseline. The fair reductions constrained
regression comparison estimators had either extremely low sensitivity or
accuracy even when they improved unfairness, which is inappropriate for
this application \citep{tisdale_cost-effectiveness_2022}.

\section{DISCUSSION}\label{sec-discussion}

We proposed a new penalized fair regression method that jointly
considers multiple groups with an efficient, flexible implementation.
Our method robustly achieved a fairness-accuracy frontier beyond that of
comparison fair regression methods across numerous simulation settings.
We applied the method to develop a classifier for ESRD with a preferable
balance of accuracy, sensitivity, and true positive rate fairness across
racial and ethnic groups compared to existing methods. While the best
penalized fair regression method depends on the underlying (typically
unknown) data structure and application, our framework is well suited to
a priori specified benchmarks for selecting among the synthesis
mechanisms and fairness weights. It can seamlessly be incorporated into
ensembles.

Although this framework is general, we do not empirically demonstrate
how it performs for all fairness definitions, estimator types, or
outcomes. We also found that for small groups, our proposed method and
preexisting methods have high variance in fairness and accuracy
measures. When group sizes vary greatly, the fair estimator's accuracy
and unfairness for small groups can vary substantially based on the
choice of multi-group fairness synthesis function. In our application,
there is no universal standard for race measurement in the AFC, and we
lacked the sample size to study fairness for some racial groups,
including American Indian and Alaska Native patients. Furthermore, fair
regression is just one aspect of ethical machine learning: problem
selection, data collection, outcome definition, and postdeployment
considerations also impact fairness \citep{chen_ethical_2021}.

In the future, our framework could be adapted to other fairness
definitions, estimator types, and outcomes. Further work should explore
methods for improving fairness for small groups, including different
methods synthesizing unfairness measures across groups. Additional areas
of future work include combining fair penalties for multiple groups with
methods for adjusting for outcome imbalance, such as reweighting or
upsampling. Assessing and increasing the generalizability or
transportability of the fairness gains produced by this method may also
be an area for further development. A potential extension is
implementing a Bayesian optimization framework for penalty weights,
using the score functions to determine a posterior distribution for the
weights rather than a single optimal set \citep{snoek_practical_2012}.

We made several novel contributions to the literature beyond the
inherent novelty of a penalized regression that is fair with respect to
multiple groups. In adapting the reductions method for penalized
regression, we detailed efficient implementation practices for penalized
fair regressions. The automatic penalty selection process we developed
includes a novel score function that rescales and synthesizes accuracy
and fairness metrics. The use of a random search for penalty weights is
novel in this literature, and we characterized our implementation
details and the random search performance under different
hyperparameters.

\subsubsection*{Acknowledgements}\label{acknowledgements}
\addcontentsline{toc}{subsubsection}{Acknowledgements}

Data were accessed using the Stanford Center for Population Health
Sciences Data Core. The PHS Data Core is supported by an NIH National
Center for Advancing Translational Science Clinical and Translational
Science Award (UL1TR003142) and from internal Stanford funding. The
content is solely the responsibility of the authors and does not
necessarily represent the official views of the NIH. We thank the ABFM
PRIME Registry participating clinicians and the ABFM, without whom the
AFC would not be possible. We also thank the Health Policy and Data
Science Lab, especially Malcolm Barrett.

This project was supported by NIH grant R01LM013989, NSF grant
DGE-2146755, and AHRQ grant 5T32HS026128-07. The authors have no
conflicts of interest to disclose. It falls under a Stanford IRB, which
approved human subjects involvement and waived consent.

\subsubsection*{Supplementary Material}\label{supplementary-material}
\addcontentsline{toc}{subsubsection}{Supplementary Material}

Web Appendices, Tables, and Figures referenced in Sections 3-5 are
available later in this document. Code is available on GitHub:
\href{https://github.com/StanfordHPDS/penalized_multiple_groups}{github.com/StanfordHPDS/penalized\_multiple\_groups}.

\subsubsection*{Data availability
statement}\label{data-availability-statement}
\addcontentsline{toc}{subsubsection}{Data availability statement}

The AFC data that support the findings in this paper cannot be shared
publicly per the data use agreement.

\newpage{}

\clearpage

\begin{table}

\caption{\label{tbl-baselogreg}\textbf{Unpenalized Regression Performance by Race and Ethnicity Group.}}

\centering{

\begin{tabular}{@{} l *{5}{S[table-format=3.2]} @{}}
\toprule
 & {n} & {Prevalence} & {Sensitivity} & {Specificity} & {Accuracy} \\
\midrule
Black & 8178 & 0.07 & 0.38 & 0.95 & 0.92 \\
Hispanic non-white & 2695 & 0.07 & 0.40 & 0.95 & 0.91 \\
Asian & 1434 & 0.07 & 0.19 & 0.98 & 0.92 \\
White & 53793 & 0.04 & 0.38 & 0.97 & 0.95 \\
Total & 67901 & 0.04 & 0.37 & 0.97 & 0.95 \\
\bottomrule
\end{tabular}

}

\end{table}%

\clearpage

\begin{table}

\caption{\label{tbl-esrdresmain}\textbf{Performance for Baseline Unpenalized Regression, Novel Penalized Regressions, and Fair Reductions Methods.}}

\centering{

\sisetup{detect-weight=true,detect-inline-weight=math, text-series-to-math = true, propagate-math-font = true}
\newrobustcmd\B{\DeclareFontSeriesDefault[rm]{bf}{b}\bfseries}
\addtolength{\jot}{-1.5em}
\begin{tabular}{@{} l *{6}{S[table-format=3.2]} @{}}
\toprule
\multicolumn{3}{r}{} & \multicolumn{4}{c}{Unfairness} \\
\cline{4-7}{Method} & {Accuracy} & {Sensitivity} & {Mean} & {Black} & {Asian} & {HNW} \\
\midrule
Baseline & 0.95 & 0.38 & 0.07 & 0.06 & 0.15 & 0.08 \\
\makecell[l]{$\begin{aligned} \text{Pop } \alpha=\,&0.1, 0.2, 0.3, 0.4 \\ &0.5, 0.6, 0.7, 0.8, 0.9, 0.99 \end{aligned}$} & 0.95 & 0.39 & 0.01 & 0.00 & 0.10 & 0.00 \\
\makecell[l]{Max $\alpha=0.2, 0.3$\\$\begin{aligned} \text{Group } \alpha=\,&0.1, 0.2, 0.3, 0.4 \\ &0.5, 0.6 \end{aligned}$} & 0.95 & 0.32 & 0.04 & 0.04 & 0.04 & 0.02 \\
\makecell[l]{Max $\alpha=0.1$} & 0.95 & 0.27 & 0.03 & 0.04 & 0.02 & 0.01 \\
\makecell[l]{$\begin{aligned} \text{Max } \alpha=\,&0.4, 0.5, 0.6, 0.7 \\ &0.8, 0.9, 0.99 \end{aligned}$\\[-4pt]Group $\alpha=0.7, 0.8, 0.9, 0.99$} & 0.90 & 0.31 & 0.05 & 0.06 & 0.00 & 0.03 \\
ExpGrad, Ep = 0.1 & 0.96 & 0.14 & 0.00 & 0.00 & 0.00 & 0.00 \\
ExpGrad, Ep = 0.01 & 0.95 & 0.04 & 0.03 & 0.04 & 0.01 & 0.00 \\
Grid, CW = 0.25 & 0.40 & 0.83 & 0.02 & 0.00 & 0.17 & 0.00 \\
Grid, CW = 0.5 & 0.40 & 0.83 & 0.02 & 0.00 & 0.17 & 0.00 \\
Grid, CW = 0.75 & 0.40 & 0.83 & 0.02 & 0.00 & 0.17 & 0.00 \\
\bottomrule
\end{tabular}
\vspace{0.2cm} 

 \begin{flushleft} \small{\textbf{Note:} Novel penalized fair regressions were defined based on the multi-group unfairness synthesis mechanism: population-weighted average (Pop), group-weighted average (Group), and maximum (Max) as well as fairness score weight ($\alpha$). Fair reductions comparisons were exponentiated gradient (ExpGrad) or grid search (Grid) methods, varying constraint flexibility (Ep) or constraint weight (CW), respectively. `HNW' is Hispanic non-white.} \end{flushleft}

}

\end{table}%

\clearpage

\section*{Supplementary Material}\label{supplementary-material-1}
\addcontentsline{toc}{section}{Supplementary Material}

\section*{Web Appendix A: Continuous
Outcomes}\label{web-appendix-a-continuous-outcomes}
\addcontentsline{toc}{section}{Web Appendix A: Continuous Outcomes}

Here we seek to estimate continuous outcome \(Y\). The baseline
continuous outcome regression generates estimates \(\hat Y\) as a
function of \(m\) input variables \(X\), indexed by \(q\), by
identifying the estimator that minimizes the mean squared error (MSE)
loss. The unfairness metric is mean residual disparity, defined as the
difference in mean residuals between the reference group and the group
of interest: \[
U_{MR} = \frac{1}{n_r} \sum_{k \in r} (\hat Y_k - Y_k) - \frac{1}{n_g} \sum_{j\in g} (\hat Y_j - Y_j).
\]

Penalized regression with continuous outcomes will add penalties for
unfairness as defined above, where each group has its own penalty term
and penalty weight.

\subsection*{Example: Lasso}\label{example-lasso}
\addcontentsline{toc}{subsection}{Example: Lasso}

We first use the example of lasso, which involves adding a penalty on
regression coefficients \(\beta\) with penalty weight \(\gamma\) to the
mean squared error loss and estimating \(\hat Y\) as a linear
combination of regression coefficients and predictors: \[
L = \sum_{i=1}^n \left[ \sum_{q=1}^m \beta_q X_{iq} - Y_i \right]^2 + \gamma \sum_{q=1}^m |\beta_q|.
\]

The loss function with fairness penalties is given by: \[
L = \sum_{i=1}^n \left[ \sum_{q=1}^m \beta_q X_{iq} - Y_i \right]^2 + \gamma \sum_{q=1}^m |\beta_q| + \sum_{g \in G} \lambda_g \left[\frac{1}{n_r} \sum_{k \in r} (\hat Y_k - Y_k) - \frac{1}{n_g} \sum_{j\in g} (\hat Y_j - Y_j) \right].
\]

This loss function minimization problem can be solved using the L-BFGS
algorithm, a type of quasi-Newtonian optimizer \citep{liu_limited_1989},
available in the \texttt{PyTorch} package.

\subsection*{Example: Neural Net}\label{example-neural-net}
\addcontentsline{toc}{subsection}{Example: Neural Net}

A penalized loss function for a simple neural network with two hidden
layers could be defined as:

\[
L = \sum_{i=1}^n \left[ \sum_{q=1}^m \beta_q X_{iq} - Y_i \right]^2 + \sum_{g \in G} \lambda_g \left[\frac{1}{n_r} \sum_{k \in r} (\hat Y_k - Y_k) - \frac{1}{n_g} \sum_{j\in g} (\hat Y_j - Y_j) \right].
\]

The loss function minimization problem here can be solved using the Adam
algorithm, a type of gradient descent-based optimizer
\citep{loshchilov_decoupled_2019}, also available in the
\texttt{PyTorch} package.

\clearpage

\section*{Web Appendix B: Penalized
Regression}\label{web-appendix-b-penalized-regression}
\addcontentsline{toc}{section}{Web Appendix B: Penalized Regression}

Additional information about how the penalized fair regression with true
positive rate disparity unfairness measures was conducted. This includes
the detailed derivation of the classifier weights for the reductions
approach and an algorithm.

\subsection*{Weighted Classification
Derivation}\label{weighted-classification-derivation}
\addcontentsline{toc}{subsection}{Weighted Classification Derivation}

We begin with the loss function from Section 3.1 of the manuscript: \[
L = - \sum_{i=1}^n \{ Y_i\ln(\hat P_i) + (1-Y_i)\ln(1 - \hat P_i) \} + \sum_{g \in G} \lambda_g n_g \left(\frac{\sum_{k\in r}Y_k\hat P_k}{\sum_{k\in r}Y_k} - \frac{\sum_{j\in g}Y_j\hat P_j}{\sum_{j\in g}Y_j} \right). 
\] We could calculate the gradient and use a gradient descent-based
algorithm to find the solution. With
\(\nabla \hat P = \hat P(1-\hat P)X\), the gradient is \[
\nabla L = - \sum_{i=1}^n \left(Y_i - \hat P_i \right)X_i + \sum_{g \in G} \lambda_g n_g \left(\frac{\sum_{k\in r} \nabla \hat P_k}{\sum_{k\in r}Y_k} - \frac{\sum_{j\in g}Y_j \nabla \hat P_j}{\sum_{j\in g}Y_j} \right) .
\]

However, we use a more efficient method based on the reductions approach
of \citet{agarwal_reductions_2018}. First, we reconceptualize the
problem as cost minimization loss, which punishes misalignment between
\(\hat P_i\) and \(Y_i\) with a cost
\citep{zadrozny_cost-sensitive_2003}. The cost-sensitive loss function
with the standard erroneous classification costs of 1 for all
observations has the same solution as the standard binary cross entropy
loss. We add this cost-sensitive loss to the true positive rate
disparity unfairness penalty terms to get the cost minimization
penalized loss function: \[
L = \sum_{i=1}^n \left\{\hat P_i (1 - Y_i) + (1 - \hat P_i) Y_i \right\} + \sum_{g \in G} \lambda_g n_g \left(\frac{\sum_{k\in r}Y_k\hat P_k}{\sum_{k\in r}Y_k} - \frac{\sum_{j\in g}Y_j\hat P_j}{\sum_{j\in g}Y_j} \right) .
\] This penalized loss function is, itself, a cost-sensitive
classification loss function with more complex costs. This is apparent
when the function is rewritten: \[
L = \sum_{i=1}^n \left[\hat P_i \left\{1 - Y_i + Y_i \sum_{g \in G}\lambda_g n_g \left( \frac{\mathbb{I}_{ri}}{\sum_{k\in r}Y_k } - \frac{\mathbb{I}_{gi}}{\sum_{j\in g}Y_j} \right) \right\} + (1 - \hat P_i) Y_i   \right]. 
\]

Reconceptualizing of the penalized fair regression as a cost-sensitive
classification problem, the costs for positive (\(C_i^1\)) and negative
(\(C_i^0\)) classifications are
\[ C_i^1 = 1 - Y_i + Y_i \sum_{g \in G}\lambda_g n_g \left( \frac{\mathbb{I}_{ri}}{\sum_{k\in r}Y_k } - \frac{\mathbb{I}_{gi}}{\sum_{j\in g}Y_j} \right), \]
and \(C_i^0 = Y_i\). These costs differ from the standard cost of 1 for
all incorrect classifications and 0 for all correct classifications. For
reference group individuals, even correct positive classifications incur
some cost, and for individuals in the groups \(g \in G\), correct
positive classifications will incur negative costs, which could also be
called rewards.

With this formulation, we extend a theorem from
\citet{zadrozny_cost-sensitive_2003} to rewrite the cost-sensitive
classification problem as a weighted non-cost-sensitive classification
problem. This yields the observation weights and modified outcomes
described in Section 3.1, where the weights are
\begin{equation}\phantomsection\label{eq-classwts}{
W_i = |C_i^0 - C_i^1| = \left|Y_i\left \{ 2 - \sum_{g \in G}\lambda_g n_g \left( \frac{\mathbb{I}_{ri}}{\sum_{k\in r}Y_k } - \frac{\mathbb{I}_{gi}}{\sum_{j\in g}Y_j}  \right) \right\} - 1\right|,
}\end{equation} and modified outcomes are
\(Y_i' = \mathbb{I}(C_i^0 > C_i^1)\). These weights and modified
outcomes can be used with an off-the-shelf classifier package to fit the
fair penalized estimator.

\subsection*{Penalized Loss Algorithm}\label{penalized-loss-algorithm}
\addcontentsline{toc}{subsection}{Penalized Loss Algorithm}

The algorithm uses Equation \eqref{eq-classwts} to calculate classifier
weights. With accuracy
\(A = \frac1n \sum_{i=1}^n \mathbb{I}(Y_i= \hat Y_{i})\) and superscript
\(UP\) denoting results from an unpenalized estimator, the score
function is as follows:

\begin{equation}\phantomsection\label{eq-score}{
S = \frac{A^{UP} - A}{A^{UP} - 0.5} + \frac{\sum_{g \in G}n_g \max \left( \frac{\sum_{k\in r}Y_k\hat Y_k}{\sum_{k\in r}Y_k} - \frac{\sum_{j\in g}Y_j \hat Y_{j}}{\sum_{j\in g}Y_j}, 0\right)}{\sum_{g \in G}n_g \max \left(\frac{\sum_{k\in r}Y_k\hat Y^{UP}_k}{\sum_{k\in r}Y_k} - \frac{\sum_{j\in g}Y_j\hat Y^{UP}_j}{\sum_{j\in g}Y_j},0\right)}
}\end{equation}

With these two equations defined, the algorithm is as follows:

\begin{algorithm}[h]
    \begin{algorithmic}[1]
        \REQUIRE Predictors $X$, outcome $Y$, groups $g \in G$
        \ENSURE Penalized fair regression estimator
        \FOR{$t = 1:T$ with $T$ total penalty weight set draws}
            \STATE Draw penalty weights from specified distributions $\lambda_g \sim D_g \forall g \in G$
            \FOR{$v = 1:V$ with $V$ total cross validation folds}
            \STATE Calculate $W$ for training data with Equation \eqref{eq-classwts} using $Y, g, \lambda_g \forall g \in G$
            \STATE Fit estimator as weighted logistic regression using $W, X, Y$ for training data
            \STATE Calculate score function on test data using Equation \eqref{eq-score}
            \ENDFOR
          \STATE Calculate cross-validated score to select the optimal $\lambda_g \forall g \in G$
        \ENDFOR
        \STATE Calculate $W$ for all data with Equation \eqref{eq-classwts} using $Y, g,$ and the score-maximizing $\lambda_g \forall g \in G$
        \STATE Fit estimator as weighted logistic regression using $W, X, Y$ for all data
        \RETURN Penalized fair regression estimator
    \end{algorithmic}
    \caption{Penalized Fair Regression for Multiple Groups}
    \label{alg:my-algorithm}
\end{algorithm}

\clearpage

\section*{Web Appendix C: Simulation
Study}\label{web-appendix-c-simulation-study}
\addcontentsline{toc}{section}{Web Appendix C: Simulation Study}

\subsection*{Synthetic Data
Distribution}\label{synthetic-data-distribution}
\addcontentsline{toc}{subsection}{Synthetic Data Distribution}

Let \(g_A\), \(g_B\), and \(g_C\) be the three different groups whose
membership was reflected by indicators \(\mathbb{I}_{g_A}\),
\(\mathbb{I}_{g_B}\), and \(\mathbb{I}_{g_C}\), respectively. We defined
the reference group to be the set of individuals who were not in any of
the three groups. The data were generated as follows with
\(n = 1,000,000\): \[X_1, X_2 \sim \mathcal{N} (30, 15)\]
\[X_3, X_4 \sim Pois (15)\]
\[X_5, \dots, X_9 \sim Bern(p), p = [0.2, 0.4, 0.6, 0.8, 0.95]\]
\[\mathbb{I}_{g_A} \sim Bern(\min[\max\{c_0 (X_9 + 0.1) + c_3 X_6 + \mathcal{N} (0, 0.02), 0\}, 1])\]
\[\mathbb{I}_{g_B} \sim Bern(\min[\max\{c_1 (X_7 + 0.05) + c_4 X_6 + \mathcal{N} (0, 0.02), 0\}, 1])\]
\[\mathbb{I}_{g_C} \sim Bern(\min[\max\{c_2 (X_8 + 0.4) + c_5 X_6 + \mathcal{N} (0, 0.02), 0\}, 1])\]
\[ \begin{aligned} Y &\sim Bern\Bigg(\Phi^{-1}\bigg[\frac{X_1}{30}  - \frac{ X_2}{15} + \frac{3 X_3}{50} - \frac{ X_4}{25} + \frac{1}{100}\Big\{2^{X_5 + X_6} + 6 X_7 \\ & + 10 X_8 X_9 + 8 X_3 \mathbb{I}_{g_A} + 6 X_4 \mathbb{I}_{g_B} + \mathbb{I}_{g_C} \mathcal{N} (60, 2)\Big \} + \mathcal{N} (0, 1)\bigg ]\Bigg). \end{aligned} \]

Across three simulation settings, \(c=c_0, \dots, c_5\) vary. Simulation
setting 1 has all moderately sized groups with
\(c = (0.2,0.2,0.2,0.02,0.02,0.02)\), \(g_A\) is small in simulation
setting 2 with \(c = (0.02,0.2,0.2,0.002,0.02,0.02)\), and all groups
are small in simulation setting 3 with
\(c = (0.02,0.02,0.02,0.002,0.002,0.002)\). There is substantial overlap
between the groups \(g_A\), \(g_B\), and \(g_C\) in all settings.

\subsection*{Runtime}\label{runtime}
\addcontentsline{toc}{subsection}{Runtime}

Our method, with 40 candidate penalty weight draws and 30 score function
variations, took 46 minutes to run, averaged across all simulations, as
compared to 9 minutes for the comparison Agarwal et al (2018) method, on
Stanford's research computing cluster using one 8 GB core.

\newpage{}

\begin{table}
\caption{\textbf{Group Size in Percents by Simulation Setting.} Means and
standard deviations (SD) are given across simulations.}\tabularnewline

\centering
\begin{tabular}{lccc}
\toprule
 & Group A & Group B & Group C \\
\midrule
Setting 1 & 32 (0.15) & 20 (0.14) & 37 (0.16) \\
Setting 2 & 1 (0.03) & 20 (0.11) & 37 (0.13) \\
Setting 3 & 2 (0.04) & 2 (0.04) & 3 (0.05) \\
\bottomrule
\end{tabular}
\end{table}

\newpage{}

\newpage{}

\newpage{}

\newpage{}

\begin{figure}

\begin{minipage}{0.47\linewidth}

\centering{

\pandocbounded{\includegraphics[keepaspectratio]{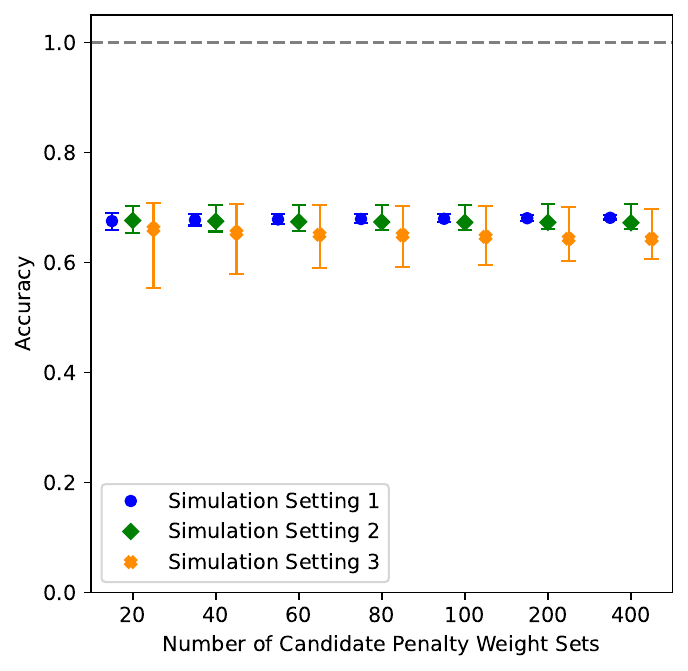}}

}

\subcaption{\label{fig-6-1}Accuracy}

\end{minipage}%
\begin{minipage}{0.06\linewidth}
~\end{minipage}%
\begin{minipage}{0.47\linewidth}

\centering{

\pandocbounded{\includegraphics[keepaspectratio]{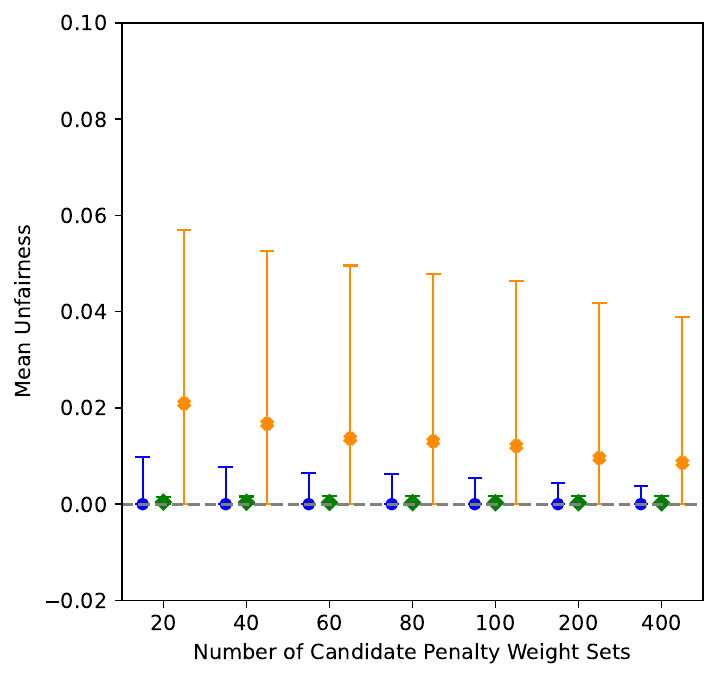}}

}

\subcaption{\label{fig-6-2}Population-Weighted Unfairness}

\end{minipage}%

\caption{\label{fig-6}\textbf{Fair Regression Estimator Accuracy and Unfairness by Number of Candidate Unfairness Penalty Weight Draws Across Simulation Settings.}
Bars represent the empirical 25th and 75th percentiles of replications.
Simulations with 40 draws of candidate penalty weight sets are
represented.}

\end{figure}%

\newpage{}

\newpage{}

\begin{figure}

\begin{minipage}{0.30\linewidth}

\centering{

\pandocbounded{\includegraphics[keepaspectratio]{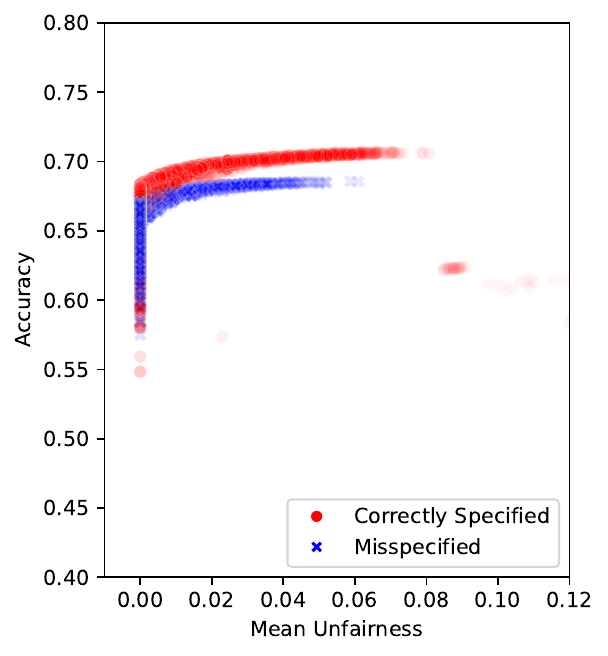}}

}

\subcaption{\label{fig-5-1}Simulation Setting 1}

\end{minipage}%
\begin{minipage}{0.05\linewidth}
~\end{minipage}%
\begin{minipage}{0.30\linewidth}

\centering{

\pandocbounded{\includegraphics[keepaspectratio]{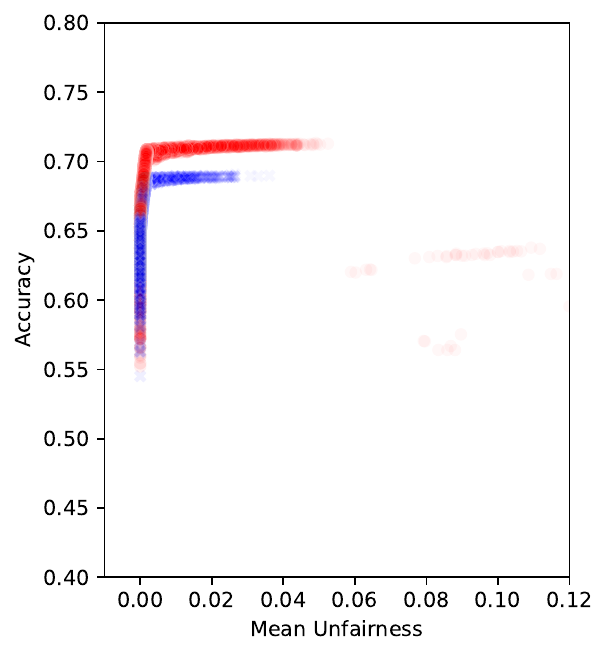}}

}

\subcaption{\label{fig-5-2}Simulation Setting 2}

\end{minipage}%
\begin{minipage}{0.05\linewidth}
~\end{minipage}%
\begin{minipage}{0.30\linewidth}

\centering{

\pandocbounded{\includegraphics[keepaspectratio]{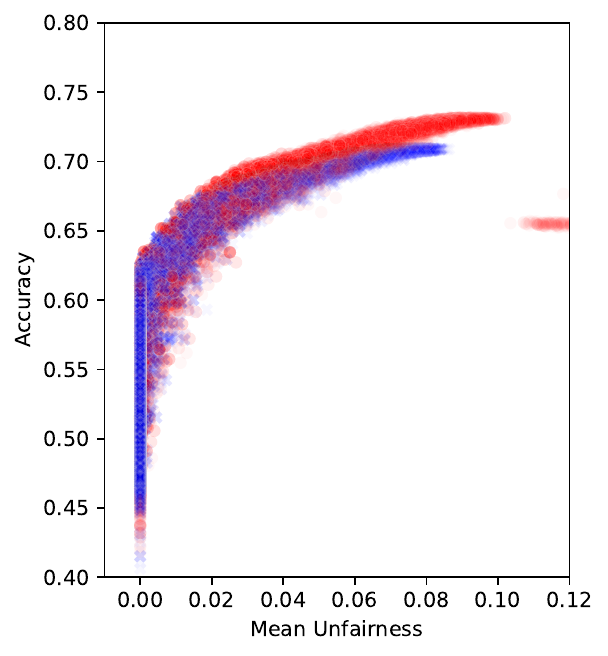}}

}

\subcaption{\label{fig-5-3}Simulation Setting 3}

\end{minipage}%

\caption{\label{fig-5}\textbf{Fair Regression Estimator Accuracy and Fairness for Correctly Specified and Misspecified Fair Regression Estimators by Simulation Setting.}
Each plotted point corresponds to the result of a single simulation
under one of the score function variations. Simulations with 40 draws of
candidate penalty weight sets are represented.}

\end{figure}%

\newpage{}

\newpage{}

\begin{figure}

\begin{minipage}{0.47\linewidth}

\centering{

\pandocbounded{\includegraphics[keepaspectratio]{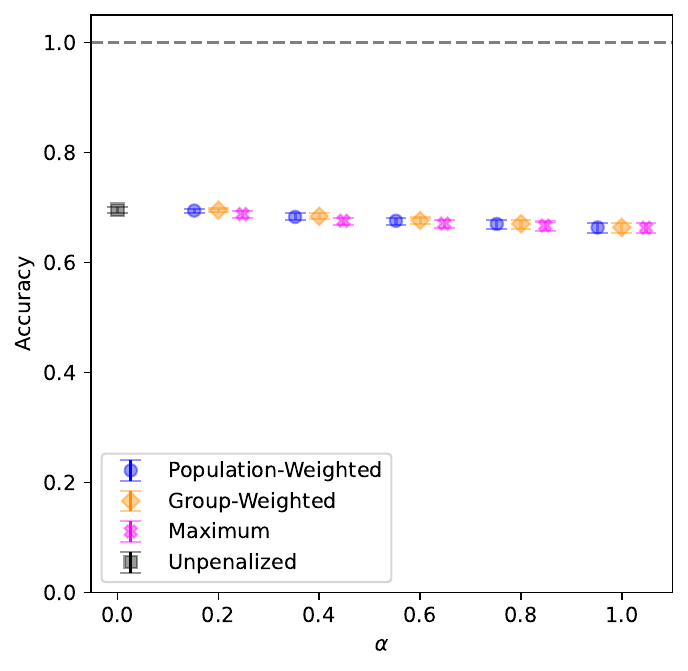}}

}

\subcaption{\label{fig-7-1}Accuracy}

\end{minipage}%
\begin{minipage}{0.06\linewidth}
~\end{minipage}%
\begin{minipage}{0.47\linewidth}

\centering{

\pandocbounded{\includegraphics[keepaspectratio]{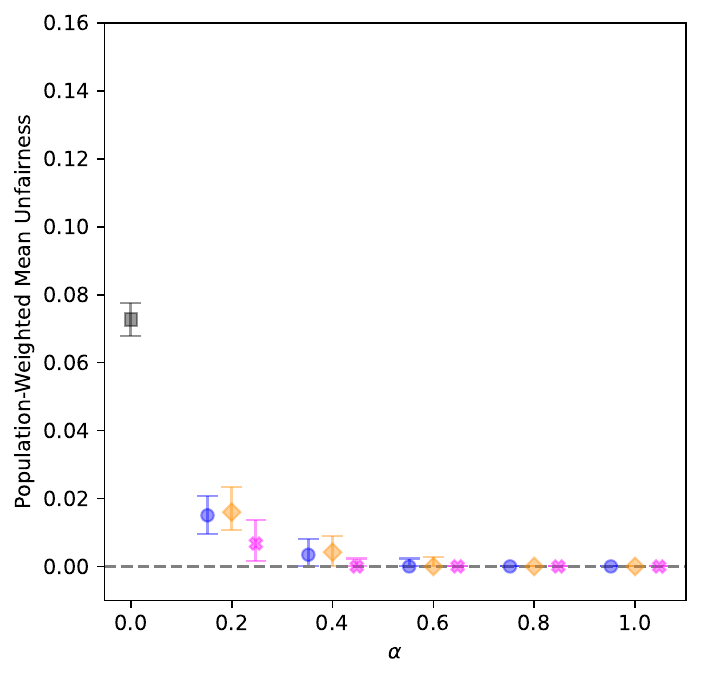}}

}

\subcaption{\label{fig-7-2}Population-Weighted Unfairness}

\end{minipage}%
\newline
\begin{minipage}{0.47\linewidth}

\centering{

\pandocbounded{\includegraphics[keepaspectratio]{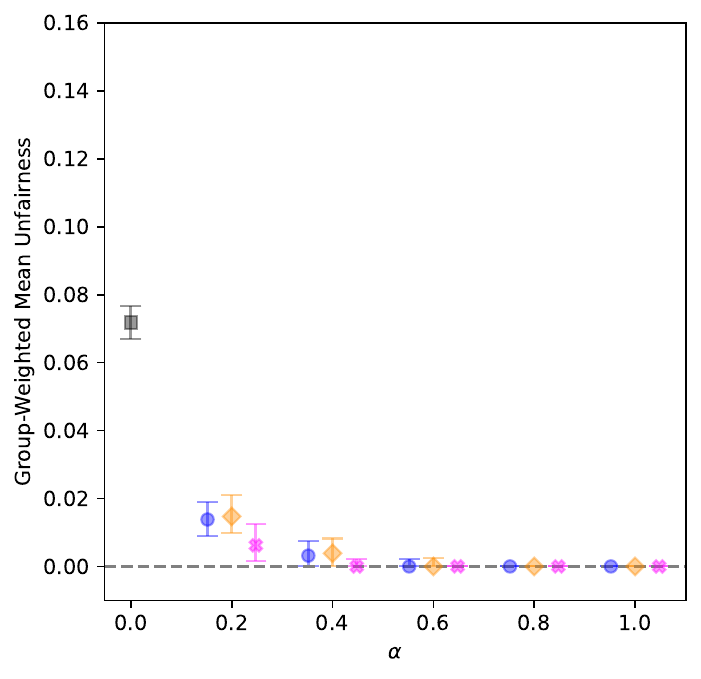}}

}

\subcaption{\label{fig-7-3}}

\end{minipage}%
\begin{minipage}{0.06\linewidth}
~\end{minipage}%
\begin{minipage}{0.47\linewidth}

\centering{

\pandocbounded{\includegraphics[keepaspectratio]{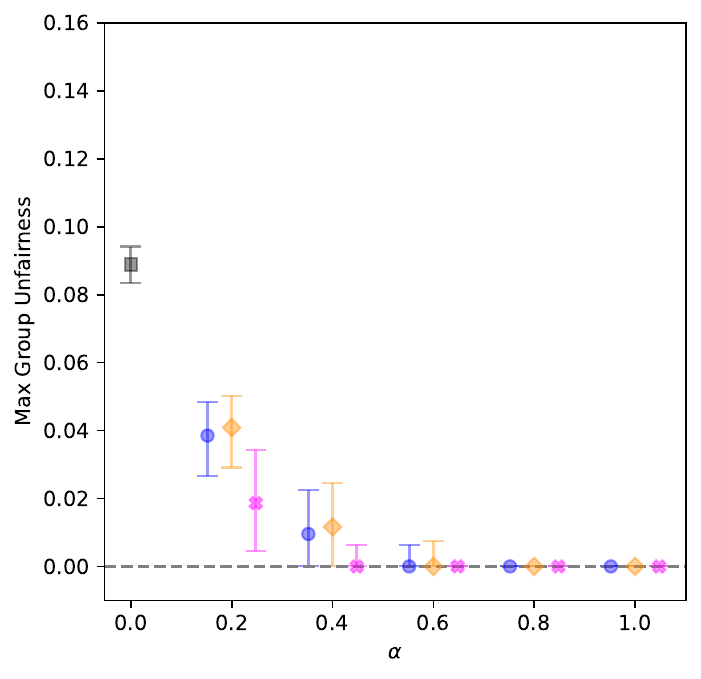}}

}

\subcaption{\label{fig-7-4}}

\end{minipage}%

\caption{\label{fig-7}\textbf{Fair Regression Estimator Accuracy and Unfairness for Different Fairness Penalty Score Weights by Multi-Group Unfairness Synthesis Function for Simulation Setting 1.}
Bars represent the empirical 25th and 75th percentiles of replications.
Simulations with 40 draws of candidate penalty weight sets are
represented.}

\end{figure}%

\newpage{}

\begin{figure}

\begin{minipage}{0.47\linewidth}

\centering{

\pandocbounded{\includegraphics[keepaspectratio]{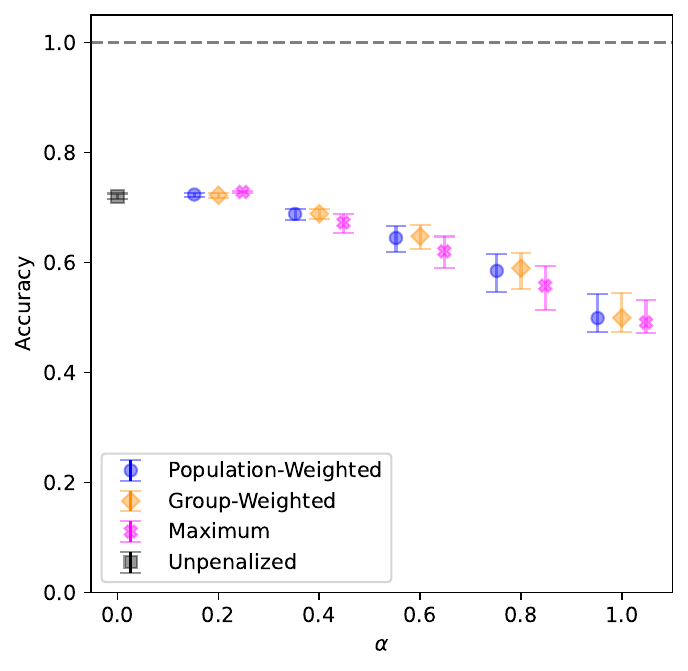}}

}

\subcaption{\label{fig-8-1}Accuracy}

\end{minipage}%
\begin{minipage}{0.06\linewidth}
~\end{minipage}%
\begin{minipage}{0.47\linewidth}

\centering{

\pandocbounded{\includegraphics[keepaspectratio]{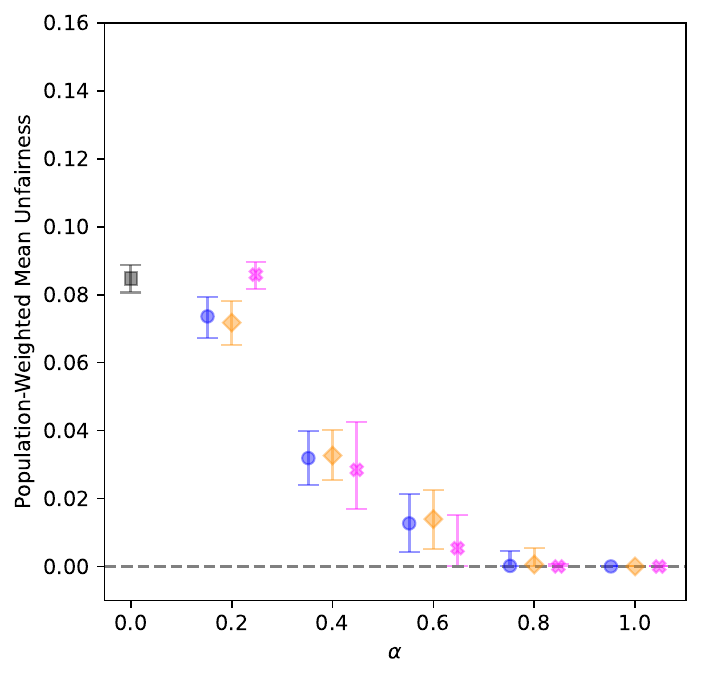}}

}

\subcaption{\label{fig-8-2}Population-Weighted Unfairness}

\end{minipage}%
\newline
\begin{minipage}{0.47\linewidth}

\centering{

\pandocbounded{\includegraphics[keepaspectratio]{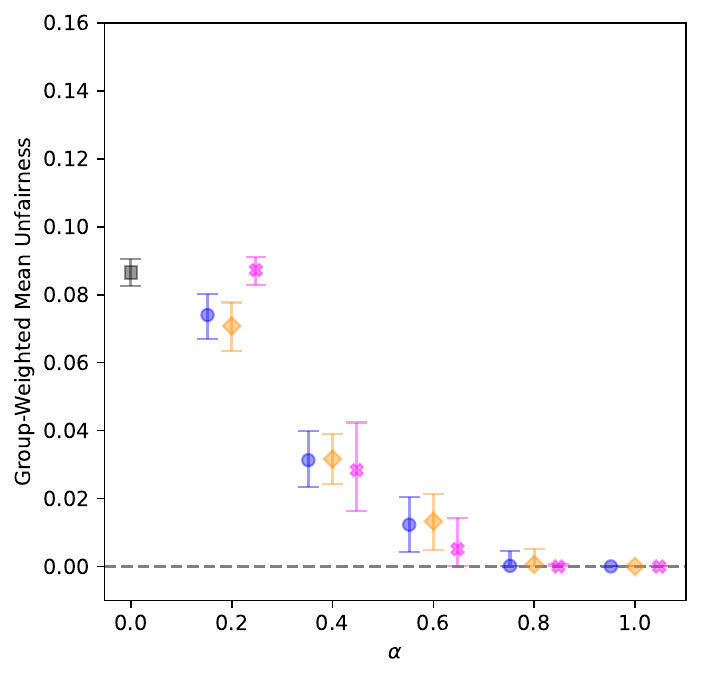}}

}

\subcaption{\label{fig-8-3}}

\end{minipage}%
\begin{minipage}{0.06\linewidth}
~\end{minipage}%
\begin{minipage}{0.47\linewidth}

\centering{

\pandocbounded{\includegraphics[keepaspectratio]{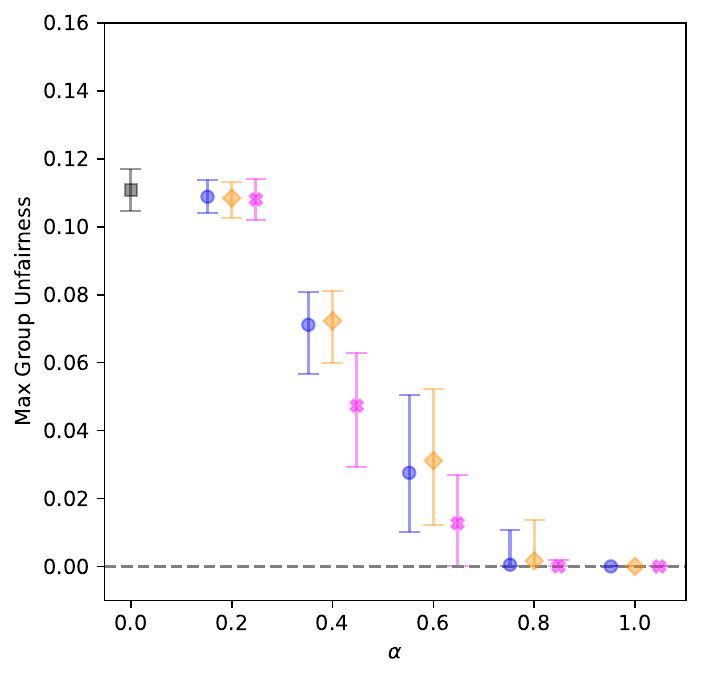}}

}

\subcaption{\label{fig-8-4}}

\end{minipage}%

\caption{\label{fig-8}\textbf{Fair Regression Estimator Accuracy and Unfairness for Different Fairness Penalty Score Weights by Multi-Group Unfairness Synthesis Function for Simulation Setting 3.}
Bars represent the empirical 25th and 75th percentiles of replications.
Simulations with 40 draws of candidate penalty weight sets are
represented.}

\end{figure}%

\clearpage

\section*{Web Appendix D: Error Rate
Disparity}\label{web-appendix-d-error-rate-disparity}
\addcontentsline{toc}{section}{Web Appendix D: Error Rate Disparity}

\subsection*{Methods}\label{methods}
\addcontentsline{toc}{subsection}{Methods}

The unfairness measure can also be based on error rate disparity. The
error rate disparity measure for a given group \(g\) is \[
U_{ER} = \frac{1}{n_g} \sum_{j \in g} (\hat Y_j (1 -Y_j) + (1 - \hat Y_j)Y_j) - \frac{1}{n_r} \sum_{k \in r} (\hat Y_k (1 -Y_k) + (1 - \hat Y_k)Y_k).
\]

As we do in the main analyses for true positive rate disparity, we will
work with an analog of this measure where we replace the predicted
binary values with predicted probabilities: \[
U_{ERP} = \frac{1}{n_g} \sum_{j \in g} (\hat P_j (1 -Y_j) + (1 - \hat P_j)Y_j) - \frac{1}{n_r} \sum_{k \in r} (\hat P_k (1 -Y_k) + (1 - \hat P_k)Y_k).
\]

The loss function with fairness penalties will have a cross-entropy loss
term, as in the primary true positive rate specification, but replaces
the true positive rate unfairness penalty terms with error rate
disparity penalty terms:
\[ \begin{aligned} L &= - \sum_{i=1}^n \{ Y_i\ln(\hat P_i) + (1-Y_i)\ln(1 - \hat P_i) \} \\ &+ \sum_{g \in G} \lambda_g n_g \left(\frac{1}{n_g} \sum_{j \in g} (\hat P_j (1 -Y_j) + (1 - \hat P_j)Y_j) - \frac{1}{n_r} \sum_{k \in r} (\hat P_k (1 -Y_k) + (1 - \hat P_k)Y_k) \right). \end{aligned} \]

We can then apply the reductions method described previously to this
loss function. The cost-sensitive loss function with the standard
erroneous classification costs of 1 for all observations, which has the
same solution as the standard binary cross entropy loss, can be added to
the true positive rate disparity unfairness penalty terms to yield the
cost minimization penalized loss function:
\[ \begin{aligned} L &= \sum_{i=1}^n \left\{\hat P_i (1 - Y_i) + (1 - \hat P_i) Y_i \right\} \\ &+ \sum_{g \in G} \lambda_g \left(\sum_{j \in g} (\hat P_g (1 -Y_g) + (1 - \hat P_g)Y_g) - \frac{n_g}{n_r} \sum_{k \in r} (\hat P_k (1 -Y_k) + (1 - \hat P_k)Y_k) \right). \end{aligned} \]

This penalized loss function is, itself, a cost-sensitive classification
loss function with more complex costs. To find the costs, we rewrite
this equation in terms of
\(J_i = \hat P_i (1 - Y_i) + (1 - \hat P_i) Y_i\) and group and
reference group indicators \(\mathbb{I}_{gi}\) and \(\mathbb{I}_{ri}\):
\[L = \sum_{i=1}^n J_i \left(1 + \sum_{g \in G} \lambda_g \left(\mathbb{I}_{gi} - \frac{n_g}{n_r}\mathbb{I}_{ri}\right) \right). \]

In this case, the costs for all errors are increased if the individual
belongs to one or more groups and decreased if they are part of the
reference group. The costs for positive (\(C_i^1\)) and negative
(\(C_i^0\)) classifications are
\(C_i^0 = Y_i \left (1 + \sum_{g \in G} \lambda_g \left(\mathbb{I}_{gi} - \frac{n_g}{n_r}\mathbb{I}_{ri}\right) \right)\)
and
\(C_i^1 = (1 - Y_i) \left(1 + \sum_{g \in G} \lambda_g \left(\mathbb{I}_{gi} - \frac{n_g}{n_r}\mathbb{I}_{ri}\right) \right)\).

With this formulation, we extend a theorem from
\citet{zadrozny_cost-sensitive_2003} to rewrite the cost-sensitive
classification problem as a weighted non-cost-sensitive classification
problem. In this formulation, the weights are \[
W_i = |C_i^0 - C_i^1| = \left|(2Y_i - 1) \left(1 + \sum_{g \in G} \lambda_g \left(\mathbb{I}_{gi} - \frac{n_g}{n_r}\mathbb{I}_{ri}\right) \right) \right|,
\] and modified outcomes are \(Y_i' = \mathbb{I}(C_i^0 > C_i^1)\). As
with the true positive rate penality, a variety of classifiers can be
use as long as they allow individual-level sample weighting. We used
random forest classifiers as implemented by \texttt{scikit-learn} with
default hyperparameters.

The score function is also revised for the error rate disparity metric.
We still use \(-s = \alpha U_{RS} + (1-\alpha) A_{RS}\), but with \[
U_{RS} = 
\frac{\sum_{g \in G}n_g \max \left( U_{ER}^g, 0\right)}{\sum_{g \in G}n_g \max \left(U_{ER,UP}^g,0\right)}.
\]

The random search process is otherwise identical to the primary
analyses.

\newpage{}

\newpage{}

\begin{figure}

\begin{minipage}{0.47\linewidth}

\centering{

\pandocbounded{\includegraphics[keepaspectratio]{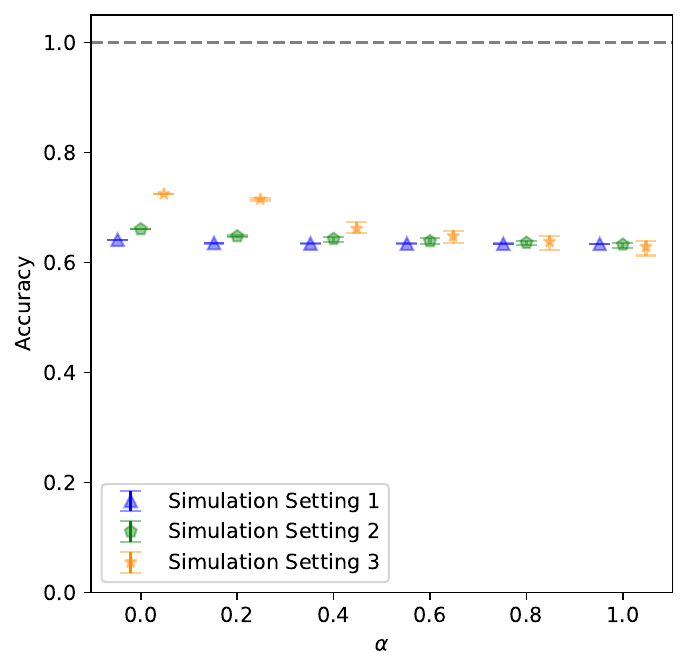}}

}

\subcaption{\label{fig-2b-1}Accuracy}

\end{minipage}%
\begin{minipage}{0.06\linewidth}
~\end{minipage}%
\begin{minipage}{0.47\linewidth}

\centering{

\pandocbounded{\includegraphics[keepaspectratio]{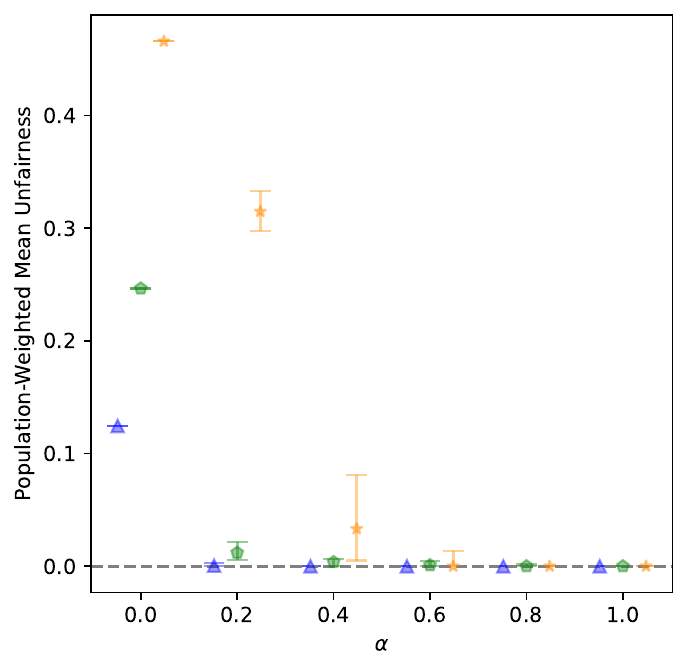}}

}

\subcaption{\label{fig-2b-2}Population-Weighted Mean Unfairness}

\end{minipage}%
\newline
\begin{minipage}{0.47\linewidth}

\centering{

\pandocbounded{\includegraphics[keepaspectratio]{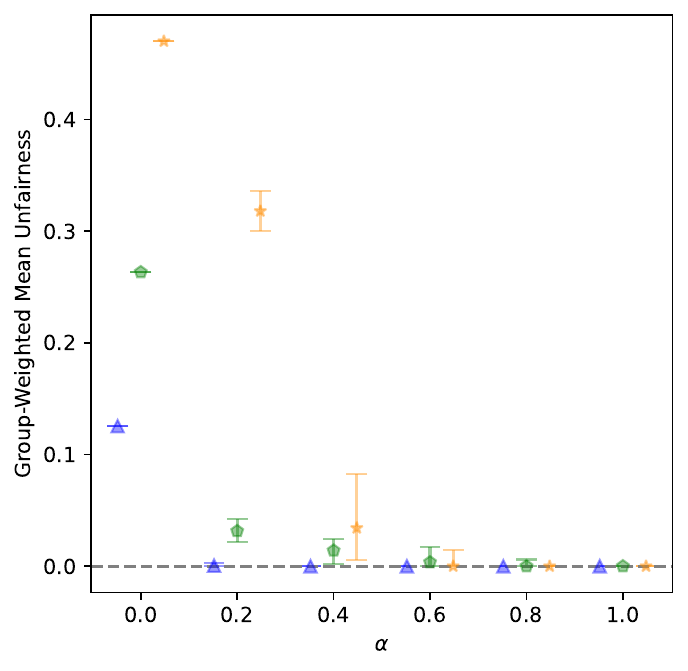}}

}

\subcaption{\label{fig-2b-3}Group-Weighted Mean Unfairness}

\end{minipage}%
\begin{minipage}{0.06\linewidth}
~\end{minipage}%
\begin{minipage}{0.47\linewidth}

\centering{

\pandocbounded{\includegraphics[keepaspectratio]{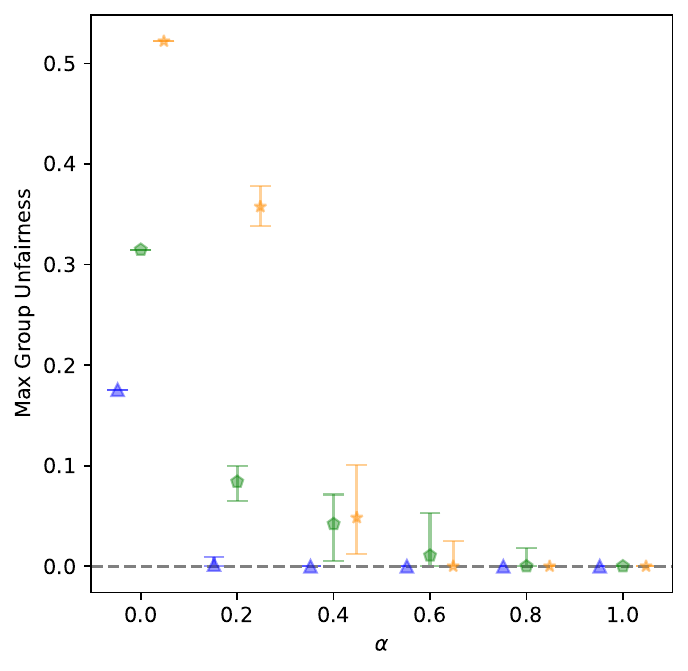}}

}

\subcaption{\label{fig-2b-4}Max Group Unfairness}

\end{minipage}%

\caption{\label{fig-2b}\textbf{Fair Regression Estimator Accuracy and Unfairness for Different Fairness Score Weights by Simulation Setting (Error Rate Disparity Fairness Measure with Random Forest Estimator).}
Bars represent the empirical 25th and 75th percentiles of replications.
Simulations with 40 draws of candidate penalty weight sets are
represented.}

\end{figure}%

\newpage{}

\begin{figure}

\begin{minipage}{0.30\linewidth}

\centering{

\pandocbounded{\includegraphics[keepaspectratio]{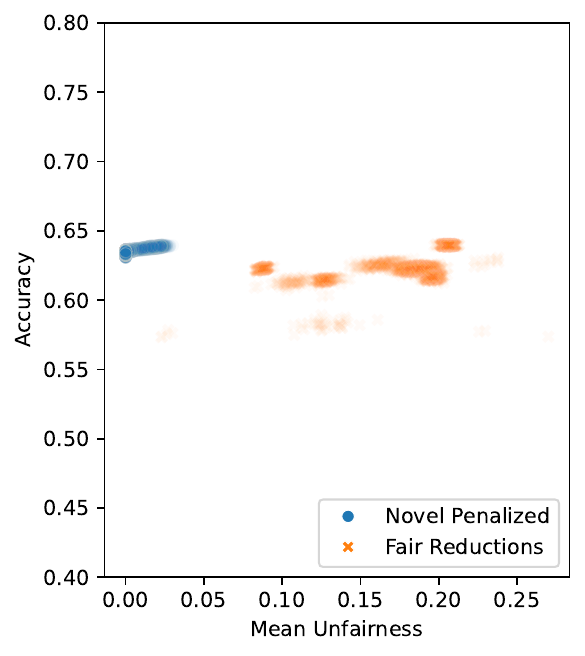}}

}

\subcaption{\label{fig-4b-1}Simulation Setting 1}

\end{minipage}%
\begin{minipage}{0.05\linewidth}
~\end{minipage}%
\begin{minipage}{0.30\linewidth}

\centering{

\pandocbounded{\includegraphics[keepaspectratio]{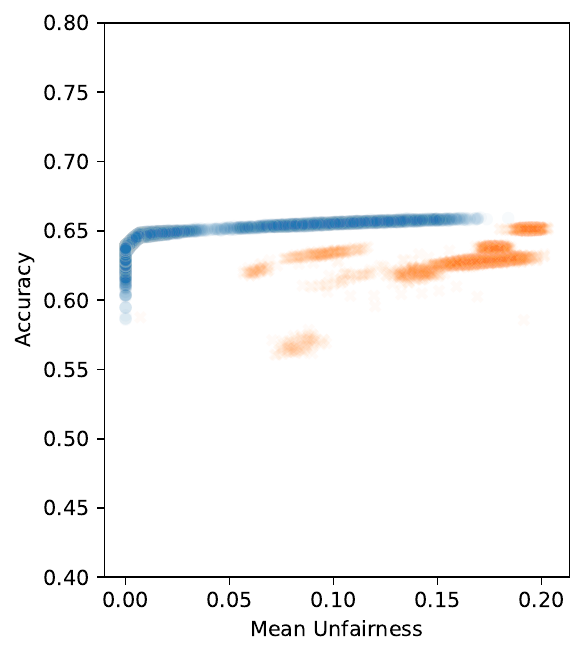}}

}

\subcaption{\label{fig-4b-2}Simulation Setting 2}

\end{minipage}%
\begin{minipage}{0.05\linewidth}
~\end{minipage}%
\begin{minipage}{0.30\linewidth}

\centering{

\pandocbounded{\includegraphics[keepaspectratio]{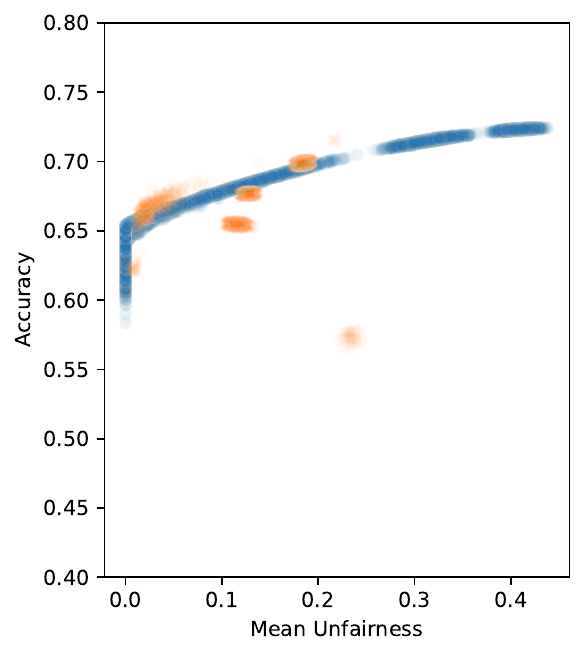}}

}

\subcaption{\label{fig-4b-3}Simulation Setting 3}

\end{minipage}%

\caption{\label{fig-4b}\textbf{Accuracy and Mean Fairness for Proposed and Comparison Fair Reductions Methods by Simulation Setting (Error Rate Disparity Fairness Measure with Random Forest Estimator).}
Each plotted point corresponds to the result of a single simulation
under one of the score function variations or enumerated fair reductions
hyperparameters. Simulations with 40 draws of candidate penalty weight
sets are represented.}

\end{figure}%

\newpage{}

\begin{figure}

\begin{minipage}{0.47\linewidth}

\centering{

\pandocbounded{\includegraphics[keepaspectratio]{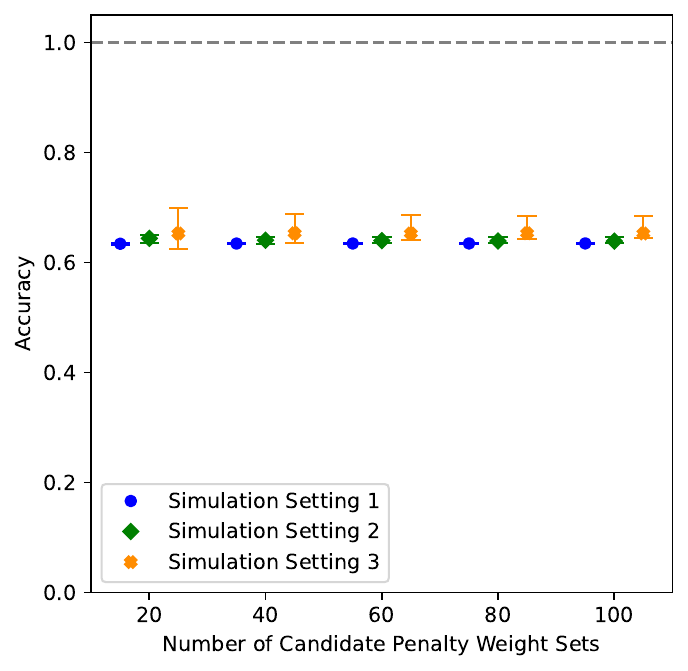}}

}

\subcaption{\label{fig-6b-1}Accuracy}

\end{minipage}%
\begin{minipage}{0.06\linewidth}
~\end{minipage}%
\begin{minipage}{0.47\linewidth}

\centering{

\pandocbounded{\includegraphics[keepaspectratio]{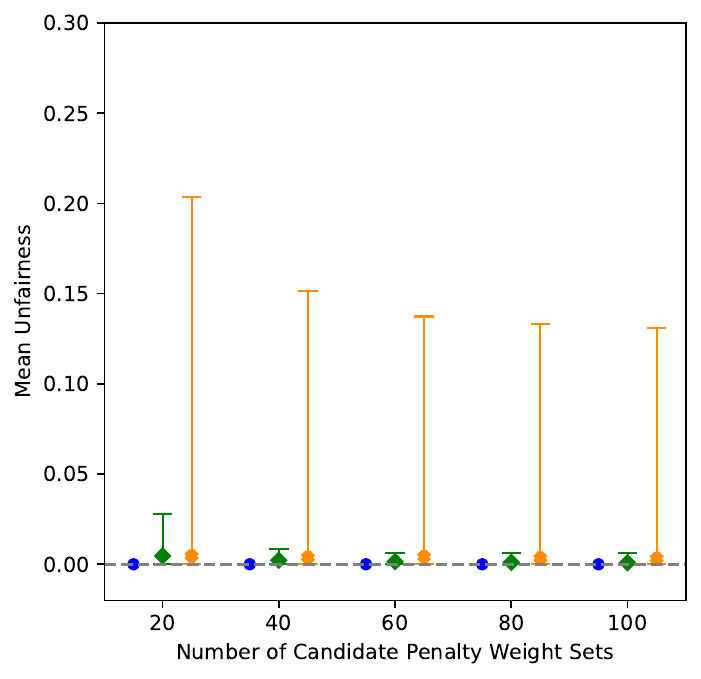}}

}

\subcaption{\label{fig-6b-2}Population-Weighted Unfairness}

\end{minipage}%

\caption{\label{fig-6b}\textbf{Fair Regression Estimator Accuracy and Unfairness by Number of Candidate Unfairness Penalty Weight Draws Across Simulation Settings (Error Rate Disparity Fairness Measure with Random Forest Estimator).}
Bars represent the empirical 25th and 75th percentiles of replications.}

\end{figure}%

\clearpage

\section*{Web Appendix E: CKD Application
Details}\label{web-appendix-e-ckd-application-details}
\addcontentsline{toc}{section}{Web Appendix E: CKD Application Details}

\begin{longtable}[]{@{}
  >{\raggedright\arraybackslash}p{(\linewidth - 4\tabcolsep) * \real{0.2558}}
  >{\raggedright\arraybackslash}p{(\linewidth - 4\tabcolsep) * \real{0.2791}}
  >{\raggedright\arraybackslash}p{(\linewidth - 4\tabcolsep) * \real{0.4651}}@{}}
\caption{\textbf{Identification Strategy for ABFM Variables.} For most
variables, we queried for the enumerated OMOP concepts in the enumerated
fields. For other variables, we queried the OMOP concept lookup table
for the OMOP ABFM data with the given text.}\tabularnewline
\toprule\noalign{}
\begin{minipage}[b]{\linewidth}\raggedright
Diagnosis
\end{minipage} & \begin{minipage}[b]{\linewidth}\raggedright
Search Method
\end{minipage} & \begin{minipage}[b]{\linewidth}\raggedright
Query
\end{minipage} \\
\midrule\noalign{}
\endfirsthead
\toprule\noalign{}
\begin{minipage}[b]{\linewidth}\raggedright
Diagnosis
\end{minipage} & \begin{minipage}[b]{\linewidth}\raggedright
Search Method
\end{minipage} & \begin{minipage}[b]{\linewidth}\raggedright
Query
\end{minipage} \\
\midrule\noalign{}
\endhead
\bottomrule\noalign{}
\endlastfoot
ESRD & Text Query & ``End stage renal disease'' or ``End stage renal
failure'' \\
eGFR & Preexisting & See
\href{https://github.com/StanfordHPDS/data_transformation/tree/main/egfr_microsim/cohort/sql_scripts}{repository} \\
HbA1c & OMOP Concepts: Measurement & 4184637, 36304734 \\
Bilirubin & OMOP Concepts: Measurement & 4118986, 3024148, 4216632,
3024128 \\
Diastolic BP & OMOP Concepts: Measurement & 4298393, 3012888 \\
Systolic BP & OMOP Concepts: Measurement & 4152194, 3004249 \\
CKD & OMOP Concepts: Condition & 44784621, 443611, 443601, 44782690,
43021852, 45768812, 44782728, 45763855, 45763854, 43531578, 45768812,
443612, 44782429, 43531653, 443614, 43021852, 443597 \\
Diabetes (Type 2) & OMOP Concepts: Condition / ICD10 & 44809548,
37208172 / E11 \\
Kidney Stone & OMOP Concepts: Condition & 201620, 45548655, 44822046 \\
Obstruction & OMOP Concepts: Condition & 45612141, 45592125, 45567838,
36713449, 4220439, 433813 \\
Hypertension Emergency & OMOP Concepts: Condition & 43020424, 45768449,
37200492, 37200493, 37200494 \\
Proteinuria & Text Query & ``Proteinuria'' \\
Glomerulonephritis & Text Query & ``Glomerulonephritis'' \\
Pyelonephritis & Text Query & ``Pyelonephritis'' \\
Female & OMOP Concepts: Gender & 8532 \\
White & OMOP Concepts: Race & 8527, 38003614 \\
Black & OMOP Concepts: Race & 8516, 38003597, 38003598, 38003599,
38003600 \\
Asian & OMOP Concepts: Race & 8515, 38003574-38003596 \\
Hispanic & OMOP Concepts: Ethnicity & 38003563 \\
\end{longtable}

\newpage{}

\begin{table}

\caption{\label{tbl-tab1}\textbf{Patient Characteristics in 2017-2018 by ESRD Status in 2019.}}

\centering{

\begin{tabular}{lllll}
\toprule
 & Missing & Overall & No ESRD & ESRD \\
\midrule
n & 0 & 67901 & 65084 & 2817 \\
Measurements &  &  &  &  \\
Minimum eGFR, mean (SD) & 37179 & 48.3 (18.7) & 49.5 (17.9) & 22.3 (16.0) \\
Mean Diastolic BP, mean (SD) & 0 & 75.3 (7.9) & 75.4 (7.9) & 74.2 (8.8) \\
Mean Systolic BP, mean (SD) & 0 & 136.6 (11.9) & 136.5 (11.7) & 140.0 (14.1) \\
Mean HbA1c, mean (SD) & 44687 & 6.8 (1.4) & 6.8 (1.4) & 6.9 (1.6) \\
Mean Bilirubin, mean (SD) & 34732 & 0.6 (0.6) & 0.6 (0.6) & 0.5 (0.3) \\
Diagnoses &  &  &  &  \\
Diabetes (Type 2), n (\%) & 0 & 109 (0.2) & 91 (0.1) & 18 (0.6) \\
Kidney Stone, n (\%) & 0 & 496 (0.7) & 453 (0.7) & 43 (1.5) \\
Proteinuria, n (\%) & 0 & 249 (0.4) & 230 (0.4) & 19 (0.7) \\
Pyelonephritis, n (\%) & 0 & 128 (0.2) & 105 (0.2) & 23 (0.8) \\
Demographics &  &  &  &  \\
Age, mean (SD) & 0 & 75.0 (11.3) & 75.1 (11.3) & 72.0 (12.3) \\
Female, n (\%) & 0 & 35926 (52.9) & 34546 (53.1) & 1380 (49.0) \\
White, n (\%) & 0 & 53793 (79.2) & 51902 (79.7) & 1891 (67.1) \\
Black, n (\%) & 0 & 8178 (12.0) & 7639 (11.7) & 539 (19.1) \\
Asian, n (\%) & 0 & 1434 (2.1) & 1330 (2.0) & 104 (3.7) \\
Hispanic non-white, n (\%) & 0 & 32927 (48.5) & 31132 (47.8) & 1795 (63.7) \\
\bottomrule
\end{tabular}

}

\end{table}%

\clearpage

\begin{figure}

\centering{

\pandocbounded{\includegraphics[keepaspectratio]{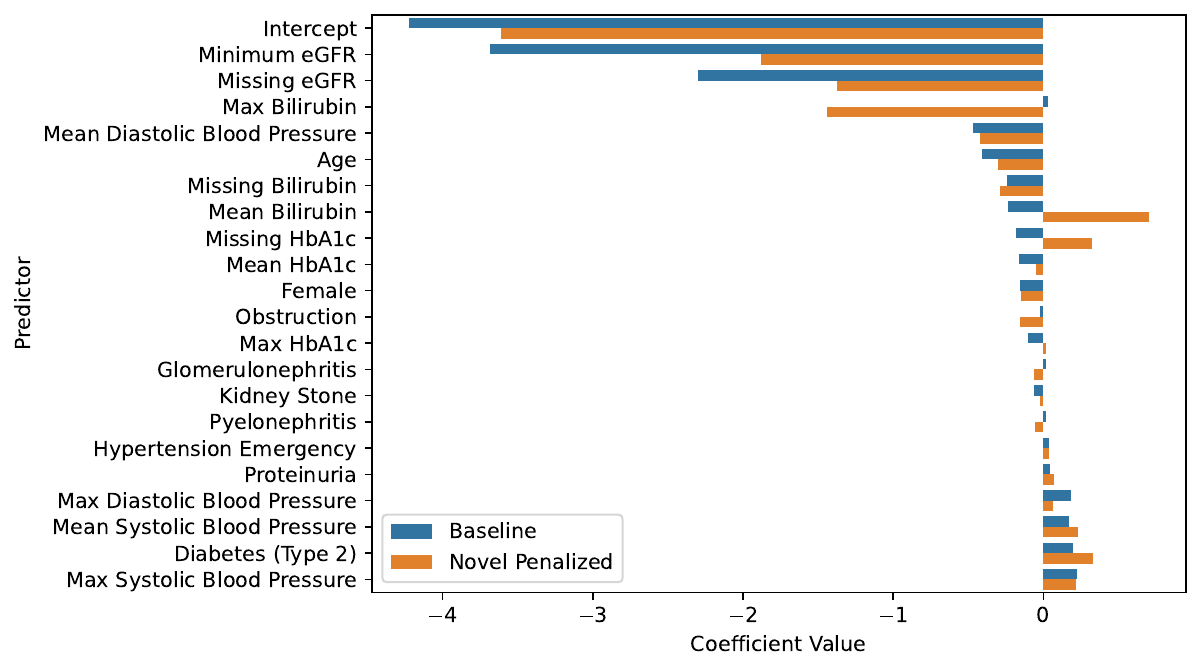}}

}

\caption{\label{fig-coefs}\textbf{Coefficients for Baseline Unpenalized and Novel Penalized Regressions.}
Predictors have been standardized to facilitate comparison of
coefficient magnitudes.}

\end{figure}%

\newpage{}

\subsection*{References}\label{references}
\addcontentsline{toc}{subsection}{References}

\renewcommand{\bibsection}{}
\bibliography{fairpenalties.bib}

\end{document}